\documentclass[12pt]{article}
\usepackage{epsfig}
\usepackage{amsfonts}
\usepackage{amsmath}
\usepackage[T1]{fontenc}
\usepackage[latin2]{inputenc}
\def\bea{\begin{eqnarray}}
\def\eea{\end{eqnarray}}
\def\bq{\begin{quote}}
\def\eq{\end{quote}}

\def\gappeq{\mathrel{\rlap
{\raise.5ex\hbox{$>$}}
{\lower.5ex\hbox{$\sim$}}}}
\def\lappeq{\mathrel{\rlap{\raise.5ex\hbox{$<$}}
{\lower.5ex\hbox{$\sim$}}}}

\def\simgt{\stackrel{>}{{}_\sim}}
\newcommand{\beq}{\begin{equation}}
\newcommand{\eeq}{\end{equation}}
\def\qo{Q_\sigma}
\def\ds{\delta s}
\def\odot{\dot{\sigma}}

\def\mP{{M_P}}
\def\xikt{\xi}

\begin{document}

\title{\bf Curvature and isocurvature perturbations \\ in two-field inflation }
\author{Z.~Lalak$^1$, D.~Langlois$^{2,3}$, S.~Pokorski$^1$, K.~Turzy\'nski$^{1,4}$\\
{\small {}}\\
{\small ${}^1${\it Institute of Theoretical Physics, Warsaw University,}}\\
{\small {\it ul.\ Ho\.za 69, 00-681 Warsaw, Poland;  }}\\
{\small ${}^2${\it APC (Astroparticules et Cosmologie, CNRS-UMR 7164), Universit\'e Paris 7}}\\
{\small {\it  10, rue Alice Domon et Léonie Duquet, 75205 Paris Cedex 13, France;}}\\
{\small ${}^3${\it GReCO, Institut d'Astrophysique de Paris, CNRS,}}\\
{\small {\it 98bis Boulevard Arago, 75014 Paris, France;  }}\\
{\small ${}^4${\it Physics Department, University of Michigan,}}\\
{\small {\it 450 Church St., Ann Arbor, MI-48109, USA  }}\\
}
\date{\today}
\maketitle

\begin{abstract}
We study cosmological perturbations in two-field inflation,
 allowing for non-standard kinetic terms. We calculate analytically the spectra of curvature and isocurvature modes at Hubble crossing, up to first order in the slow-roll parameters. We 
also compute numerically the evolution of the curvature and isocurvature modes from well within the Hubble radius until the end of inflation. We show explicitly for a few examples, including the recently proposed model 
of `roulette' inflation, how isocurvature perturbations affect significantly the curvature perturbation between Hubble crossing  and the end of inflation.
\end{abstract}

\newpage

\section{Introduction}

Inflation provides a simple and elegant scenario for the early Universe (see e.g. \cite{mukhanov} for a recent textbook presentation). 
Although single field inflation models are perfectly compatible with the present cosmological data, 
many early universe models based on high energy physics, in particular  derived from supergravity or string theory, usually involve many scalar fields which can have non-standard kinetic terms. This is why multi-field inflationary scenarios, where several scalar fields play a dynamical role during inflation, have received some   attention in the literature (see e.g. \cite{Kofman:1986wm}-\cite{vanTent:2003mn}). However, except for a few specific models, the predictions 
for the spectra of primordial perturbations are, in general, a nontrivial task, in contrast with
single-field models. 

The main reason  is that the curvature 
(or adiabatic) perturbation, which is generated during inflation 
and eventually observed, can evolve on super-Hubble scales in multi-field inflation whereas it remains frozen in single-field inflation. This is due to the presence of additional perturbation modes, often called isocurvature (or entropy) modes,  corresponding to relative perturbations between the various scalar fields, which act 
as a source term in the evolution equation for the curvature perturbation. 
This property was first pointed out in \cite{sy} in the context of Brans-Dicke inflation. 
This phenomenon occurs during 
inflation and affects the final curvature perturbation at the end of inflation, independently 
whether isocurvature modes survive or not after inflation. 

The purpose of the present work is to study in detail how the isocurvature perturbations, present during 
inflation, affect the curvature perturbations, {\it both at Hubble crossing and in the subsequent evolution 
on super-Hubble scales}. Since our intention is to stress some qualitative properties specific to multi-field inflation, we have chosen to restrict our study to the case of two scalar fields. Moreover, we consider models
where a non-standard kinetic term is allowed for one of the scalar fields. This includes, in particular, 
scenarios motivated by supergravity and string theory, which have been recently proposed (see, e.g., \cite{Bond:2006nc}-\cite{deCarlos:2007dp}).

The production of adiabatic and isocurvature modes for two-field inflation with a generic potential, in the slow-roll approximation, was studied in \cite{Gordon:2000hv} where a decomposition into 
 adiabatic and isocurvature modes was introduced. Models with non-standard kinetic terms for inflatons
have been studied in the slow-roll approximation in
\cite{Mukhanov:1997fw} and \cite{Starobinsky:2001xq},
and the adiabatic-isocurvature decomposition technique of \cite{Gordon:2000hv}
was later extended 
to such two-field models in \cite{DiMarco:2002eb}
and \cite{DiMarco:2005nq}.
Very recently, two-field inflation with standard 
kinetic terms was investigated in \cite{Byrnes:2006fr}
at next-to-leading order correction in 
a slow-roll expansion and it was also shown that the  
adiabatic and isocurvature modes at Hubble crossing are correlated 
at first order in slow-roll parameters. 
In parallel to these analytical studies, a numerical study of the 
evolution of  adiabatic and isocurvature was presented 
in \cite{Tsujikawa:2002qx} and \cite{Choi:2007su}.

In the present work, we extend the previous analyses in the following 
directions. 
First, we present a detailed analysis of the correlation of 
adiabatic and isocurvature just after Hubble crossing, both analytically and
numerically. This correlation was in general supposed to vanish in most previous works, except in the numerical study \cite{Tsujikawa:2002qx} and the analytical work \cite{Byrnes:2006fr}, both in the context of canonical 
kinetic terms.  Here, we extend the analysis with non-standard kinetic terms. We compute analytically 
the spectra and correlation at Hubble crossing in the slow-roll approximation, by taking special care of the 
time-dependence shortly after Hubble crossing. 

Second, we study numerically the whole  evolution of  adiabatic and 
isocurvature perturbations  from within the Hubble radius until  the end of inflation. 
This allows us to go beyond the slow-roll approximation 
which is needed
to derive analytical results. Our numerical study enables us to see precisely how 
 isocurvature perturbations can be transferred into adiabatic
perturbations during the inflationary phase depending on 
the background trajectory in field space. We illustrate this behaviour by studying numerically 
three models. The last one is the so-called 'roulette' inflation model, which has been proposed 
recently \cite{Bond:2006nc}.

The plan of the paper is the following. The next section presents the class of models we consider and gives the homogeneous equations of motion as well as the equations governing the perturbations. The third section is devoted to the study of the perturbations from deep inside the Hubble radius  until a few e-folds after Hubble crossing. 
We then discuss, in section 4, analytical methods to determine
the evolution of the perturbations on super-Hubble scales.
 Section 5 is devoted to the numerical study of the evolution of the perturbations, which is compared with the analytical estimates of the previous sections. We finally draw our conclusions in the last section. 

\section{The model}
In this paper, we  study models with two scalar fields, in which one of the scalar fields has
a non-standard kinetic term, described by an action of the form
\beq
\label{actionnc}
S = \int d^4 x \sqrt{-g} \left[ \frac{M_P^2}{2}R - \frac{1}{2}
(\partial_\mu\phi)(\partial^\mu\phi) - \frac{e^{2b(\phi)}}{2}
(\partial_\mu\chi)(\partial^\mu\chi) - V (\phi, \chi) \right],
\eeq
where $\mP$ is the reduced Planck mass, $\mP\equiv (8\pi G)^{-1/2}$. 
This type of action usually appears when $\chi$ corresponds to an axionic component.
 It is also motivated by generalized Einstein theories \cite{sy,Starobinsky:2001xq}.
When $b(\phi)=0$ one recovers standard kinetic terms for the two fields. In this section, we
give the equations of motion for the homogeneous fields and then for the linear perturbations, following the 
results (and notation) of \cite{DiMarco:2002eb} and \cite{DiMarco:2005nq}, where the same type of models was 
considered.

\subsection{Homogeneous equations}
\label{homogeneous}
Let us start with the homogeneous equations of motion. 
We assume  a spatially flat FLRW (Friedmann-Lema\^itre-Robertson-Walker)  geometry, with metric
\beq
ds^2=-dt^2+a(t)^2 d{\bf x}^2,
\eeq
where $t$ is the cosmic time. One can also define the comoving time $\tau=\int dt/a(t)$.

The equations of motion for the scale factor and the homogeneous fields read
\beq
\ddot \phi + 3 H \dot \phi + V_\phi = b_\phi
e^{2 b} \dot \chi^2 \,,
\label{phi_b1}
\eeq
\beq
\ddot \chi + (3 H + 2 b_\phi \dot \phi) \dot \chi 
+ e^{- 2 b} \, V_\chi = 0 \,,
\label{chi_b1}
\eeq
\beq
H^2 = \frac{1}{3 M_P^2} \left[ \frac{1}{2} \dot \phi^2 + 
\frac{e^{2 b}}{2} \dot \chi^2 + V \right] \,,
\eeq
and
\beq
\dot H = - \frac{1}{2 M_P^2} \left[ \dot \phi^2 +
e^{2 b} \dot \chi^2
\right] \, ,
\eeq
where $H\equiv \dot a/a$ and 
a dot stands for a derivative with respect to the cosmic time $t$ and a subscript index $\phi$ or $\chi$
denotes a derivative with respect to the corresponding field.

It is also useful to introduce the following slow-roll parameters 
\beq
\epsilon_{\phi\phi}=\frac{\dot{\phi}^2}{2M_P^2H^2}, \quad 
\epsilon_{\phi\chi}=e^b\frac{\dot{\phi}\dot{\chi}}{2M_P^2H^2}, \quad
\epsilon_{\chi\chi}=e^{2b}\frac{\dot{\chi}^2}{2M_P^2H^2},
\eeq
\beq 
\eta_{IJ}=\frac{V_{IJ}}{3H^2}
\eeq
and 
\beq
\label{epsilon}
\epsilon=\epsilon_{\phi\phi}+\epsilon_{\chi\chi}=-\frac{\dot H}{H^2}.
\eeq

\subsection{Linear perturbations}
We now discuss the linear perturbations of our model (one can find a detailed presentation  of the 
theory of cosmological perturbations in e.g. \cite{mukhanov,ks,mfb} and a pedagogical introduction 
in e.g. \cite{cargese}). 
For simplicity, we shall directly work 
in the longitudinal gauge. In the absence of anistropic
stress (the off-diagonal spatial components of the stress-energy tensor), 
which is the case when matter consists of scalar fields,  
the metric in the longitudinal gauge is of the form 
\beq
\mathrm{d}s^2 = -( 1 + 2 \Phi) \mathrm{d}t^2 + a^2 (1 - 2 \Phi) \mathrm{d} {\bf x}^2 \, . \label{metric}
\eeq
where  only scalar perturbations are taken into account.

We now decompose the scalar fields
into their homogeneous (background) parts and the perturbations:
\beq
\phi(t,{\bf x}) = \phi(t) +\delta\phi(t,{\bf x}) \qquad \textrm{and} \qquad
\chi(t,{\bf x}) = \chi(t) +\delta\chi(t,{\bf x}) \, .
\eeq
We shall work with the Fourier components
of the perturbations, $\delta\phi_\mathbf{k}(t)$ and $\delta\chi_\mathbf{k}(t)$,
routinely omitting the subscript $\mathbf{k}$ to shorten the expressions.
The perturbed Klein-Gordon equations read
\begin{eqnarray}
\ddot{\delta \phi} &+& 3 H \dot{\delta \phi} + 
 \left[ \frac{k^2}{a^2} + V_{\phi \phi} 
- (b_{\phi \phi} + 2 b_\phi^2) \dot \chi^2 e^{2 b} \right] 
\delta \phi +
V_{\phi \chi} \delta \chi - 2 b_\phi e^{2 b} \dot \chi \dot{\delta \chi} 
\nonumber \\
&=& 4 \dot \phi \dot \Phi - 2 V_\phi \Phi \, 
\label{deltaphi}
\end{eqnarray}
and  
\begin{eqnarray}
\ddot{\delta \chi} &+& (3 H + 2 b_\phi \dot \phi) \dot{\delta \chi} + 
\left[ \frac{k^2}{a^2} + e^{- 2 b} V_{\chi \chi} \right] \delta \chi 
+ 2 b_\phi \dot \chi \, \dot{\delta \phi}
+ 
\nonumber \\
&+& \left[ e^{- 2 b} \left( V_{\chi \phi} - 2 b_\phi V_\chi\right) + 2 b_{\phi \phi} \dot \phi \dot \chi \right] \delta \phi 
= 4 \dot \chi \dot \Phi - 2 e^{- 2 b} V_\chi \Phi \, .
\label{deltachi}
\end{eqnarray}

The energy and the momentum constraints, given by Einstein equations,
are, respectively:
\beq
3 H (\dot \Phi + H \Phi) + \dot H \Phi  + \frac{k^2}{a^2} \Phi 
= - \frac{1}{2 M_P^2} 
\left[ \dot \phi\, \dot{\delta \phi} + e^{2 b} \dot \chi\,
\dot \delta \chi + b_\phi e^{2 b} \dot \chi^2 \delta \phi
+ V_\phi \delta \phi + V_\chi \delta \chi \right]
\label{energy_constraint}
\eeq
\beq
\dot \Phi + H \Phi = \frac{1}{2 M_P^2} \left( \dot \phi \,
\delta \varphi + e^{2 b} \dot \chi \,\delta \chi \right)\,\, .
\label{momentum_constraint}
\eeq

It is convenient, instead of using the perturbations $\delta\phi$ and $\delta\chi$, defined here
in the longitudinal gauge, to introduce the so-called
gauge-invariant Mukhanov-Sasaki variables:
\beq
Q_\phi \equiv \delta\phi + \frac{\dot\phi}{H}\Phi \qquad\textrm{and}\qquad 
Q_\chi \equiv \delta\chi + \frac{\dot\chi}{H}\Phi \, ,
\label{musava}
\eeq 
which can be identified with the scalar field perturbations in the flat gauge. 

Substituting (\ref{musava}) into (\ref{deltaphi})-(\ref{deltachi}) and using the
background equations of motion as well as the energy and momentum
constraints, one finds 
\begin{eqnarray}
\label{eom1c}
&& \ddot{Q}_\phi+3H\dot{Q}_\phi-2e^{2b}b_\phi\,\dot{\chi}\,\dot{Q}_\chi 
+\left(\frac{k^2}{a^2}+C_{\phi\phi}\right)Q_\phi + C_{\phi\chi}Q_\chi =0\\
\label{eom2c}
&& \ddot{Q}_\chi+3H\dot{Q}_\chi+2b_\phi\,\dot{\phi}\,\dot{Q}_\chi+2b_\phi\,\dot{\chi}\,\dot{Q}_\phi 
+\left(\frac{k^2}{a^2}+C_{\chi\chi}\right)Q_\chi + C_{\chi\phi}Q_\phi =0\, ,
\end{eqnarray}
where  we have defined the following background-dependent coefficients:
\begin{eqnarray}
C_{\phi\phi} &=&  -2e^{2b}b_\phi^2\dot{\chi}^2+\frac{3\dot{\phi}^2}{M_P^2}-\frac{e^{2b}\dot{\phi}^2\dot{\chi}^2}{2M_P^4H^2}-\frac{\dot{\phi}^4}{2M_P^4H^2}-e^{2b}b_{\phi\phi}\dot{\chi}^2+\frac{2\dot{\phi}V_{\phi}}{M_P^2H}+V_{\phi\phi} \\
C_{\phi\chi} &=& 
 \frac{3e^{2b}\dot{\phi}\dot{\chi}}{M_P^2}-\frac{e^{4b}\dot{\phi}\dot{\chi}^3}{2M_P^4H^2}-\frac{e^{2b}\dot{\phi}^3\dot{\chi}}{2M_P^4H^2}+\frac{\dot{\phi}V_\chi}{M_P^2H}+\frac{e^{2b}\dot{\chi}V_\phi}{M_P^2H}+V_{\phi\chi} \\
C_{\chi\chi} &=& 
\frac{3e^{2b}\dot{\chi}^2}{M_P^2}-\frac{e^{4b}\dot{\chi}^4}{2M_P^4H^2}-\frac{e^{2b}\dot{\phi}^2\dot{\chi}^2}{2M_P^4H^2}+\frac{2\dot{\chi}V_\chi}{M_P^2H}+e^{-2b}V_{\chi\chi} \\
C_{\chi\phi} &=&
\frac{3\dot{\phi}\dot{\chi}}{M_P^2}-\frac{e^{2b}\dot{\phi}\dot{\chi}^3}{2M_P^4H^2}-\frac{\dot{\phi}^3\dot{\chi}}{2M_P^4H^2}+2b_{\phi\phi}\dot{\phi}\dot{\chi}-2e^{-2b}b_\phi V_\chi + \nonumber \\
&& + \frac{e^{-2b}\dot{\phi}V_\chi}{M_P^2H}+\frac{\dot{\chi}V_\phi}{M_P^2H}+e^{-2b}V_{\phi\chi} 
\end{eqnarray}
The two equations (\ref{eom1c}) and (\ref{eom2c}) form a closed system for the two gauge-invariant quantities $Q_\phi$ and 
$Q_\chi$.

\subsection{Decomposition into adiabatic and entropy components}
As originally proposed in \cite{Gordon:2000hv}, 
in order to facilitate the interpretation of the evolution of cosmological perturbations, it can be useful  to decompose the scalar field perturbations along the two directions respectively parallel and orthogonal to the homogeneous trajectory in field space. The projection parallel to the trajectory is usually called 
the adiabatic, or curvature, component while the orthogonal projection corresponds to the   
 entropy, or isocurvature, component. Note that there was a semantic shift in the  terminology since one 
used to call adiabatic and entropy modes during inflation the two particular solutions for the perturbations
that would match after inflation, respectively, to the adiabatic and isocurvature modes defined in the radiation 
era. This terminology is used for example in the papers on double inflation such that  
\cite{Polarski:1992dq} and \cite{Langlois:1999dw}.

This decomposition into {\it instantaneous} adiabatic and entropy components, introduced in 
\cite{Gordon:2000hv}, has  recently been extended 
\cite{Langlois:2006vv} to fully  nonlinear perturbations in the context of the covariant nonlinear formalism introduced  
in \cite{Langlois:2005ii,Langlois:2005qp}. Here, we consider only the decomposition at the linear level, but since we allow
for non-standard kinetic terms, we will need to generalize the equations to such a case, as was done  
in \cite{DiMarco:2002eb}. Let us recall here the main results.

The essential idea is to introduce the linear combinations
\beq
\delta\sigma \equiv \cos\theta\, \delta\phi + \sin\theta\,e^b\, \delta\chi  \qquad\textrm{and}\qquad 
\ds \equiv -\sin\theta\, \delta\phi + \cos\theta\,e^b\, \delta\chi \, ,
\eeq
where
\beq
\cos\theta \equiv \frac{\dot\phi}{\dot\sigma}, \qquad
\sin\theta \equiv \frac{\dot\chi\,e^b}{\dot\sigma} \qquad\textrm{with}\qquad
\dot\sigma \equiv \sqrt{\dot\phi^2+e^{2b}\dot\chi^2} \, .
\eeq
 The notations $\dot\sigma$ and $\delta\sigma$ are just used for convenience; they do not refer to any 
scalar field $\sigma$.

Instead of $\delta\sigma$, it is in fact more convenient to work directly  with the  Mukhanov-Sasaki variables and therefore 
to define 
\beq
\qo \equiv \cos\theta\, Q_\phi + \sin\theta\,e^b\, Q_\chi  \qquad\textrm{and}\qquad 
\ds \equiv -\sin\theta\, Q_\phi + \cos\theta\,e^b\, Q_\chi \, ,
\label{musava1}
\eeq
by noting that 
\beq
\qo \equiv \delta\sigma +  \frac{\dot\sigma}{H}\Phi.
\eeq
In the so-called comoving  gauge, the perturbation
$\qo$ is directly related to the  three-dimensional
curvature of the constant time space-like slices. This gives the gauge-invariant 
quantity refered to as the comoving curvature perturbation:
\beq
\mathcal{R}\equiv\frac{H}{\odot}Q_\sigma.
\label{rcal}
\eeq
The perturbation $\ds$,
called the isocurvature perturbation, is automatically gauge-invariant. It is sometimes convenient, by 
analogy with the curvature perturbation,  
to introduce a renormalized entropy perturbation which is defined  as 
\beq
\mathcal{S}\equiv\frac{H}{\odot}\delta s.
\label{scal}
\eeq
In  field space, $\qo$ corresponds
to perturbations parallel to the velocity vector $(\dot\phi,e^b\dot\chi)$,
while $\ds$ to the orthogonal ones.

Introducing the adiabatic and entropy ``vectors'' in field space, respectively 
\beq
E_\sigma^I=\left(\cos\theta, e^{-b}\sin\theta\right), \qquad
E_s^I=\left(-\sin\theta, e^{-b}\cos\theta\right), \qquad
I=\left\{\phi,\chi\right\},
\eeq
one can define various derivatives of the potential with respect to the 
adiabatic and entropy directions. Assuming an implicit summation on the indices $I$ (and $J$), 
the first order derivatives are defined as
\beq
V_\sigma=E_\sigma^IV_I, \qquad 
V_s=E_s^IV_I,
\eeq
whereas the second order derivatives are 
\beq
V_{\sigma\sigma}=E_\sigma^I E_\sigma^JV_{IJ}, \quad
V_{\sigma s}=E_\sigma^I E_s^JV_{IJ}, \quad
V_{ss}=E_s^I E_s^JV_{IJ}.
\eeq

By combining the two Klein-Gordon equations for the background fields, (\ref{phi_b1}) and (\ref{chi_b1}), one gets
the background equations of motion along the adiabatic and entropy directions, respectively,
\bea
&& \ddot \sigma+3H\dot\sigma+V_\sigma=0,\\
&& \dot\theta=-\frac{V_s}{\dot\sigma}-b_\phi\dot\sigma \sin\theta,
\label {dot_theta}
\eea
while the equations of motion for the perturbations read:
\begin{eqnarray}
\label{eom1d}
 && \ddot{Q}_\sigma+3H\dot{Q}_\sigma
+\left(\frac{k^2}{a^2}+C_{\sigma\sigma}\right)Q_\sigma +\frac{2V_s}{\odot}\dot{\ds} + C_{\sigma s}\,\delta s 
=0
\\
\label{eom2d}
 && \ddot{\ds}+3H\dot{\ds}
+\left(\frac{k^2}{a^2}+C_{ss}\right)\ds -\frac{2V_s}{\odot}\dot{Q}_\sigma +C_{s\sigma} Q_\sigma=0 \, ,
\end{eqnarray}
with
\begin{eqnarray}
\label{cd01}
C_{\sigma\sigma} &=& V_{\sigma\sigma}-\left(\frac{V_s}{\odot}\right)^2
+2\frac{\odot V_\sigma}{M_P^2 H}
+\frac{3\odot^2}{M_P^2}-\frac{\odot^4}{2M_P^4H^2}
-b_\phi\left(s_\theta^2c_\theta V_\sigma+(c_\theta^2+1)s_\theta V_s\right)
\\
C_{\sigma s} &=& 6H\frac{V_s}{\odot}+\frac{2V_\sigma V_s}{\odot^2}+2V_{\sigma s}+ \frac{\odot V_s}{M_P^2 H}
+2b_\phi(s_\theta^3V_\sigma-c_\theta^3V_s)
\\
C_{ss} &=& V_{ss} -\left(\frac{V_s}{\odot}\right)^2
+b_\phi(1+s_\theta^2)c_\theta V_\sigma +b_\phi c_\theta^2s_\theta V_s - \dot{\sigma}^2(b_{\phi\phi}+b_\phi^2)
\\
\label{cd04}
C_{s\sigma} &=&
-6H\frac{V_s}{\odot}-\frac{2V_\sigma V_s}{\odot^2}
+\frac{\odot V_s}{M_P^2 H}
\end{eqnarray}
where $s_\theta\equiv\sin\theta$ and $c_\theta\equiv\cos\theta$.

The above system consists (\ref{eom1d}-\ref{eom2d}) of two coupled second order differential equations involving 
only $\qo$ and $\ds$.  In order to relate these variables to the metric perturbation $\Phi$ defined 
 in the longitudinal gauge, it is useful to use  the Poisson-like constraint, which follows from the energy and 
 momentum constraints (\ref{energy_constraint}) and (\ref{momentum_constraint}), 
\beq
\frac{k^2}{a^2}\Phi=-\frac{1}{2M_P^2}\epsilon_m
\label{poison1}
\eeq
where $\epsilon_m$ is the comoving energy density and can be expressed as 
\beq
\epsilon_m= \dot\sigma\dot{Q}_\sigma+\left(3H+\frac{\dot H}{H}\right)
\dot\sigma Q_\sigma+V_\sigma Q_\sigma+2V_s\delta s.
\label{poison2}
\eeq

\subsection{Perturbation spectra}
The inflationary observables are customarily expressed in terms of
power spectra and correlation functions. Given their origin as
quantum fluctuations, the perturbations can be represented as random
variables. We introduce the power spectra of the adiabatic and entropy perturbations, 
respectively 
\beq
\langle {Q_\sigma}^\ast_\mathbf{k}\, {Q_\sigma}_{\mathbf{k}'}\rangle=\frac{2\pi^2}{k^3}\mathcal{P}_{Q_\sigma}(k)\delta(\mathbf{k}-\mathbf{k}'),
\quad
\langle \ds^\ast_\mathbf{k}\, \ds_{\mathbf{k}'}\rangle=
\frac{2\pi^2}{k^3}
\mathcal{P}_{\ds}(k)\delta(\mathbf{k}-\mathbf{k}'),
\eeq
as well as the correlation spectrum
\beq
\langle {Q_\sigma}^\ast_\mathbf{k}\, \ds_{\mathbf{k}'}\rangle=\frac{2\pi^2}{k^3}
\mathcal{C}_{Q_\sigma\, \ds}(k)\delta(\mathbf{k}-\mathbf{k}').
\eeq

\section{Evolution of perturbations inside the Hubble radius}
\label{inside}

In order to study the generation of perturbations from vacuum fluctuations, 
we start, for a given comoving wave number $k$, at an instant $t_i$ (or $\tau_i$) 
during inflation when the physical wave number $k/a$ 
is much bigger than the Hubble parameter $H$.
Our initial conditions are 
given, as usual,  by the Minkowski-like vacuum at $\tau_i$,
\beq
\label{init_adiab}
Q_\sigma(\tau_i)\simeq\frac{e^{-\imath k\tau_i}}{a(\tau_i)\sqrt{2k}}
\qquad\textrm{and}\qquad
\delta s(\tau_i)\simeq\frac{e^{-\imath k\tau_i}}{a(\tau_i)\sqrt{2k}}
\eeq
for initial adiabatic and isocurvature fluctuations,
respectively.
These two initial fluctuations are statistically independent because the 
corresponding equations of motion are decoupled in the limit $k\gg aH$. 

Although the adiabatic and entropy fluctuations are initially, i.e.\ deep
inside the Hubble radius, statistically independent, this is, in general, 
no longer the case at Hubble crossing.  In the context 
of two-field inflation with {\sl canonical} kinetic terms, this point has been stressed in the 
numerical analysis of \cite{Tsujikawa:2002qx} and studied  analytically in \cite{Byrnes:2006fr}. 

In the following, we  extend the analysis of  \cite{Byrnes:2006fr} to non-canonical 
kinetic terms. 

\subsection{Equations in the slow-roll approximation}

We start with the perturbations
$Q_\sigma$ and $\delta s$, whose dynamics is described by eqs.\
(\ref{eom1d}) and (\ref{eom2d}). Using the conformal time $\tau$ and introducing 
the variables 
\beq
u_\sigma=a\,Q_\sigma, \qquad u_s=a\,\delta s,
\eeq
these equations can be rewritten in the form
\begin{eqnarray}
&&u_\sigma''+ \frac{2V_s}{\odot}a u_s'+\left[k^2-\frac{a''}{a}+a^2 {C}_{\sigma\sigma}\right]u_\sigma
+\left[-\frac{2V_s}{\odot}a' +a^2 {C}_{\sigma s}\right]u_s=0,\\
&&u_s''- \frac{2V_s}{\odot}a u_\sigma'+\left[k^2-\frac{a''}{a}+a^2 {C}_{ss}\right]u_s
+\left[\frac{2V_s}{\odot}a' +a^2 {C}_{s\sigma}\right]u_\sigma=0,
\end{eqnarray}
where the four coefficients ${C}_{IJ}$ 
are given in eqs.\ (\ref{cd01})-(\ref{cd04})
and a prime denotes a derivative with respect to the conformal time $\tau$.

Let us now discuss the slow-roll approximation. The only difference with respect to the case with canonical 
kinetic terms will 
arise from  some of the terms depending on the derivatives of $b$.
In the slow-roll approximation, one  can use the relation
\beq
\frac{V_s}{\dot{\sigma}} = H\eta_{\sigma s}-b_\phi\dot{\sigma}s_\theta^3 \, ,
\label{feq}
\eeq
and the various coefficients in the above system of equations simplify to yield
\beq
\left[ \left( \frac{\mathrm{d}^2{ }}{\mathrm{d}\tau^2}+k^2-\frac{2+3\epsilon}{\tau^2}\right)\mathbf{1}+  2\mathbf{E}\frac{1}{\tau} \frac{\mathrm{d}{ }}{\mathrm{d}\tau} +\mathbf{M} \frac{1}{\tau^2}\right] \left( \begin{array}{c} u_\sigma \\ u_s \end{array}\right)=0 \,
\label{eomcol1}
\eeq
where the matrices $\mathbf{E}$ and $\mathbf{M}$ are given by
\begin{eqnarray}
\label{ema}
\mathbf{E} &=& 
\left( \begin{array}{cc} 0 & -\eta_{\sigma s} \\ \eta_{\sigma s} & 0 \end{array}\right) 
+
\left( \begin{array}{cc} 0 & \xikt s_\theta^3 \\ -\xikt s_\theta^3 & 0 \end{array}\right) 
\\
\mathbf{M} &=& 
\left( \begin{array}{cc} -6\epsilon + 3\eta_{\sigma\sigma} & 4\eta_{\sigma s} \\ 2\eta_{\sigma s} & 3\eta_{ss} \end{array}\right)+
\left( \begin{array}{cc} 3\xikt s_\theta^2c_\theta & -4\xikt s_\theta^3 \\ -2\xikt s_\theta^3 & -3\xikt c_\theta (1+s_\theta^2) \end{array}\right)
\label{tmma}
\end{eqnarray}
with
\beq
\xikt\equiv\sqrt{2}b_\phi M_P\sqrt{\epsilon}.
\eeq
In eqs.\ (\ref{ema}) and (\ref{tmma}), we have kept only the terms linear in $b_\phi$, i.e. 
proportional to $\xikt$, and neglected the terms quadratic in $b_\phi$ as well as the terms proportional 
to $b_{\phi\phi}$. We thus treat $\xikt$ on the same footing as the other slow-roll parameters. 
In order to emphasize the difference between generalized kinetic terms and canonical kinetic terms, we have however separated the terms proportional $\xikt$ from the others.

The system of equations (\ref{eomcol1}) that we have obtained is of the form
\beq
u''+2\mathbf{L}u'+\mathbf{Q}u=0 \, .
\label{matrixeq}
\eeq
The matrix coefficient for the first order time derivative is 
 $2\mathbf{L}=2\mathbf{E}/\tau$,
where $\mathbf{E}$ is an antisymmetric matrix, linear in the slow-roll
parameters. 
Let us introduce a {\it time-dependent} orthogonal matrix $\mathbf{R}$ which satisfies $\mathbf{R}'=-\mathbf{L}\mathbf{R}$. Note that this 
is possible only if $\mathbf{L}$ is an antisymmetric matrix. Reexpressing the above equation (\ref{matrixeq}) 
in terms of
 a new matrix vector $v$, defined by $u=\mathbf{R}v$, one can eliminate the terms proportional to the first order time  
derivative and we obtain the following equation
\beq
v''+\mathbf{R}^{-1}\left(-\mathbf{L}^2-\mathbf{L}'+\mathbf{Q}\right)\mathbf{R} v=0.
\eeq
At linear order in the slow-roll parameters, one  finds
\beq
-\mathbf{L}^2-\mathbf{L}'\simeq \frac{1}{\tau^2}\mathbf{E}.
\eeq
Therefore, apart from the trivial part proportional to the identity matrix,  the combination 
$-\mathbf{L}^2-\mathbf{L}'+\mathbf{Q}$ contains
\beq
\frac{1}{\tau^2}(\mathbf{E}+\mathbf{M})=\frac{3}{\tau^2}\left( \begin{array}{cc} 
-2\epsilon+\eta_{\sigma\sigma}+\xikt s_\theta^2c_\theta & 
\eta_{\sigma s} -\xikt s_\theta^3 \\ \eta_{\sigma s}-\xikt s_\theta^3 & \eta_{ss}-\xikt c_\theta (1+s_\theta^2) \end{array}\right)  \, ,
\label{mma}
\eeq
which is a symmetric matrix.

We now assume that the slow-roll parameters vary sufficiently slowly during the few e-folds when the 
given scale crosses out the Hubble radius. We thus replace the time-dependent matrix on the 
right hand side of  (\ref{mma}) by the same matrix evaluated at Hubble crossing, i.e. for $k=aH$, and the only 
remaining time dependence appears in the global coefficient $3/\tau^2$. One can now diagonalize 
this matrix by introducing the time-independent rotation matrix 
\beq
\mathbf{\tilde{R}}_*=\left(
\begin{array}{cc}
\cos\Theta_* & -\sin\Theta_* \\
\sin\Theta_* & \cos\Theta_* 
\end{array}
\right),
\eeq
so that 
\beq
\mathbf{\tilde{R}}_*^{-1}\left(\mathbf{M}+\mathbf{E}\right) \mathbf{\tilde{R}}_*=
{\rm Diag}(\tilde{\lambda}_1,\tilde{\lambda}_2).
\eeq
In particular, one can easily compute the following linear combinations, which will be useful later:
\begin{eqnarray}
\label{vp1}
&& \tilde\lambda_1+\tilde\lambda_2=3\left(\eta_{\sigma\sigma}+\eta_{ss}-2\epsilon 
-\xikt c_\theta\right), \\
&&
\label{vp2}
(\tilde\lambda_1-\tilde\lambda_2)\sin 2\Theta_*=6 \left(\eta_{\sigma s}- \xikt  s_\theta^3\right),
\\
&&
\label{vp3}
(\tilde\lambda_1-\tilde\lambda_2)\cos 2\Theta_* =3 \left(\eta_{\sigma\sigma}-\eta_{ss}-2\epsilon 
+\xikt c_\theta (1+2 s_\theta^2)\right),
\end{eqnarray}
where the right hand sides of the three above equations are evaluated at $k=aH$.

Similarly, the rotation matrix $\mathbf{R}$ is slowly varying  per efold, since $(\mathrm{d}\mathbf{R}/\mathrm{d}\,\ln a)\mathbf{R}^T=\mathbf{E}$, where $\mathbf{E}$ is 
linear in slow-roll parameters. Around Hubble crossing, one can thus replace $\mathbf{R}$ by its value $\mathbf{R}_*$ at Hubble crossing. 

By introducing 
\beq
w=\mathbf{\tilde{R}}_*^{-1}\mathbf{R}_*v,
\eeq
the system of equations can be written as two independent equations of the form 
\beq
w_A''+\left[k^2-\frac{1}{\tau^2}\left(2+3\lambda_A\right)\right]w_A=0, \qquad \left(A=1,2\right)
\label{bessel}
\eeq
with
\beq
\label{lambda} 
 \lambda_A=\epsilon-\frac{1}{3}\tilde{\lambda}_A.
\eeq
Defining
\beq
\mu_A=\sqrt{\frac{9}{4}+3\lambda_A}
\eeq
 the solution of (\ref{bessel}) 
 with the proper asymptotic behaviour can be written as 
\beq
\label{solbes}
w_A=\frac{\sqrt{\pi}}{2}e^{i(\mu_A+1/2)\pi/2}\sqrt{-\tau}H_{\mu_A}^{(1)}(-k\tau) e_A(k),
\eeq
where $H_{\mu}^{(1)}$ is  the Hankel function of the first kind of order 
$\mu$ and the $e_A$ are two independent normalized Gaussian random variables so that 
\beq
\langle e_A(k)\rangle=0, \qquad \langle e_A(k) e_B^*(k')\rangle =\delta_{AB}
\delta^{(3)}(k-k').
\eeq

Using the independence of the variables $w_1$ and $w_2$, one can express 
the correlations for the variables $u_\sigma$ and $u_s$ around Hubble crossing time as
\begin{eqnarray}
\label{bwqsigma}
a^2 \langle Q_\sigma^\dagger Q_\sigma \rangle &=& 
\cos^2\Theta_* \langle w_1^\dagger w_1 \rangle+
\sin^2\Theta_*\langle w_2^\dagger w_2 \rangle 
 \\
a^2 \langle \delta s^\dagger Q_\sigma \rangle &=& \frac{1}{2}\sin2\Theta_* \left( \langle w_1^\dagger w_1 \rangle-\langle w_2^\dagger w_2 \rangle \right) \\
a^2 \langle \delta s^\dagger \delta s \rangle &=& \sin^2\Theta \langle w_1^\dagger w_1 \rangle+
\cos^2\Theta \langle w_2^\dagger w_2 \rangle 
\label{bwdeltas}
\end{eqnarray}
where one can substitute
\beq
\langle w_A^\dagger w_A \rangle=\frac{\pi}{4}(-\tau)|H^{(1)}_{\mu_A}(-k\tau)|^2\equiv \frac{1}{2k}\frac{1}{(k\tau)^2}{\cal F}_A(-k\tau).
\eeq
This yields
\begin{eqnarray}
{\cal P}_{Q_\sigma} &=& \left(\frac{H_*}{2\pi}\right)^2(1-2\epsilon_*)\left[\cos^2\Theta_*\  {\cal F}_1(-k\tau)
+\sin^2\Theta_*\ {\cal F}_2(-k\tau)\right]
\\
{\cal C}_{Q_\sigma \delta s} &=& \left(\frac{H_*}{2\pi}\right)^2(1-2\epsilon_*)\frac{\sin2\Theta_*}{2}\  
\left[ {\cal F}_1(-k\tau)-{\cal F}_2(-k\tau)\right]
\\
{\cal P}_{\delta s}  &=& \left(\frac{H_*}{2\pi}\right)^2(1-2\epsilon_*)
\left[\sin^2\Theta_*\  {\cal F}_1(-k\tau)+\cos^2\Theta_*\ {\cal F}_2(-k\tau) \right] ,
\end{eqnarray}
where we have used
\beq
a\simeq-\frac{1+\epsilon_*}{H_*\tau}.
\eeq
At this stage, it is worth noting that our  derivation is still valid if the parameter 
$\eta_{ss}$, which corresponds to the curvature of the potential along the direction orthogonal to the field 
trajectory,
is not small. In this case, the entropy fluctuations  are effectively suppressed and only  
adiabatic fluctuations are generated. As far as the perturbations are concerned, this particular situation 
is similar to the single field case.

A further simplification occurs when $\lambda_A\ll 1$, in which case $\mu_A\simeq \frac{3}{2}+\lambda_A$. The 
functions ${\cal F}_A(x)$ can be expanded as 
\beq
{\cal F}_A(x)=\frac{\pi}{2}x^3|H_{3/2}(x)|^2\left(1+2\lambda_A f(x)\right)=(1+x^2)\left(1+2\lambda_A f(x)\right),
\eeq
with 
\beq
f(x) = \mathrm{Re}\left( \frac{1}{H^{(1)}_{3/2}(x)}\left. \frac{\mathrm{d}H^{(1)}_\mu (x)}{\mathrm{d}\mu}\right|_{\mu=3/2} \right).
\eeq

Using the relations (\ref{vp1}-\ref{vp3}) and (\ref{lambda}), 
we finally get for the curvature and entropy perturbations defined in 
(\ref{rcal}) and (\ref{scal}), the following expressions
\begin{eqnarray}
\mathcal{P}_\mathcal{R} &=& \left(\frac{H_*^2}{2\pi\dot\sigma_*}\right)^2(1+k^2\tau^2)
\left[1-2\epsilon_*+\left(6\epsilon_*-2\eta_{\sigma\sigma *}-2\xikt_* s_{\theta *}^2c_{\theta *}\right)
\,f\left(\frac{k}{aH_*}\right)\right] 
\label{pra}
\\
\mathcal{C}_\mathcal{RS} &=& \left(\frac{H_*^2}{2\pi\dot\sigma_*}\right)^2(1+k^2\tau^2)
\left( 2\xikt_* s_{\theta *}^3-2\eta_{\sigma s*} \right)\,f\left(\frac{k}{aH_*}\right) 
\label{pca}
\\
\mathcal{P}_\mathcal{S} &=& \left(\frac{H_*^2}{2\pi\dot\sigma_*}\right)^2(1+k^2\tau^2)
\left[1-2\epsilon_*+\left(2\epsilon_*-2\eta_{ss *}+2\xikt_* (1+s_{\theta *}^2)c_{\theta *}\right)\,f\left(\frac{k}{aH_*}\right)\right].
\label{psa}
\end{eqnarray}

Let us comment these results. First, one can verify that, for
canonical kinetic terms (i.e. $\xikt=0$),  we recover  the results of \cite{Byrnes:2006fr} if we replace 
the factor $(1+k^2\tau^2)$ by $1$ and the function $f(-k\tau)$ by the number $C=2-{\rm ln}2-\gamma\simeq 0.7296$, where $\gamma\simeq 0.5772$ is the Euler-Mascheroni constant. Our final expression depends explicitly on $\tau$
and allows us a more precise estimate of the spectra around the time of Hubble crossing. 
As we will see explicitly later, 
there are some inflationary scenarios where the amplitude 
of the curvature perturbation spectrum evolves very quickly after Hubble crossing and  never reaches 
its asymptotic value (corresponding to the limit $k\tau\to 0$). In these cases, one needs to evaluate more precisely the amplitude around Hubble crossing, which our more detailed formula enables to do.

Another difference with \cite{Byrnes:2006fr} is that we derived the adiabatic and isocurvature spectra by working directly with the equations for the adiabatic and entropy components, instead of working with the initial scalar fields.

\section{Evolution of perturbations on super-Hubble scales}
\label{outside}

When the isocurvature modes are suppressed, for instance if the effective mass along the isocurvature 
direction is  large with respect to the Hubble parameter  the final adiabatic 
spectrum  can be computed simply by taking the usual single-field result applied 
to the adiabatic direction: 
\beq
{\cal P}_{\cal R}^\mathrm{sf}(k)\simeq \frac{H^4}{4\pi^2\dot\sigma^2}=\frac{H^4}{8\pi^2
\mathcal{L}_\mathrm{kin}}\, ,
\label{sifi}
\eeq
where all the quantities are evaluated at  Hubble crossing.
The simplification works because, in this particular situation where isocurvature fluctuations are 
absent, the curvature perturbation 
remains frozen on super-Hubble scales.

If  isocurvature modes are present however, they will  
affect the super-Hubble  
evolution of the adiabatic perturbations because they will act as a 
source term on the right hand side of the equation governing the 
evolution of the curvature perturbation \cite{Wands:1996kb} (and \cite{Langlois:2006vv} for 
the non-linear generalisation).

In order to obtain the final power spectra and correlations, and
to compare the predictions of a multi-inflaton model
with observations, one must then  solve the coupled
system of differential equations 
(\ref{eom1d}-\ref{eom2d}).  In general, a numerical approach is necessary and will be considered in the next
section.
In some particular cases, within the slow-roll approximation, 
one can solve analytically the equations of motion on super-Hubble scales. We
now discuss these cases in the rest of this section.

Following \cite{DiMarco:2005nq},
we can then write eqs.\ (\ref{eom1d}) and (\ref{eom2d})
in the slow-roll approximation as:
\beq
\label{sreom}
\dot{\qo} \simeq AH\qo+BH\ds \qquad\textrm{and}\qquad \dot{\ds}\simeq DH\ds \, ,
\eeq
where:
\begin{eqnarray}
\label{apar}
A &=& -\eta_{\sigma\sigma}+2\epsilon-\xikt c_\theta s^2_\theta \\
\label{bpar}
B &=& -2\eta_{\sigma s}+2\xikt s^3_\theta \simeq 2\frac{\mathrm{d}\theta}{\mathrm{d}N}-2\xikt s_\theta\\
\label{spar}
D &=& -\eta_{ss}+\xikt c_\theta (1+s^2_\theta) \, .
\end{eqnarray}
Qualitatively, it is clear that if the isocurvature perturbations
do not decay very fast, there is a strong interaction between the
adiabatic and isocurvature perturbations, whenever the coefficient
$B$ becomes large, i.e.\ when the classical trajectory makes a sharp
turn in the field space or when $\xikt$ is relatively large and inflation
is driven at least partially by the field $\chi$ ($s_\theta\neq0$).
For {\it constant} $A,B,D$, the equations 
(\ref{sreom}) can be solved explicitly to give
\beq
\label{srsol}
\qo(N) \simeq e^{AN}Q_{\sigma*} + \frac{B}{D-A}(e^{DN}-e^{AN})\delta s_* \qquad\textrm{and}
\qquad \ds(N) \simeq e^{DN} \delta s_* \,
\eeq
where
$N$ stands for the number of efolds after  Hubble crossing. 
Taking into
account that $(H/\dot{\sigma})\simeq(H_*/\dot{\sigma}_*)e^{-AN}$, we
can express the power spectra and correlations as:
\begin{eqnarray}
\label{shr}
\mathcal{P}_\mathcal{R}^{(a)}(N) &\simeq& \bar{\mathcal{P}}_{\mathcal{R}*} +
\bar{\mathcal{P}}_{\mathcal{S}*}\left(\frac{B}{\gamma}\right)^2\left(e^{\gamma N}-1\right)^2
+ 2\bar{\mathcal{C}}_{\mathcal{RS}*}
\frac{B}{\gamma}\left(e^{\gamma N}-1\right) \\
\label{shc}
\mathcal{C}_\mathcal{RS}^{(a)}(N) &\simeq&
\bar{\mathcal{C}}_{\mathcal{RS}*}\,e^{\gamma N}+\bar{\mathcal{P}}_{\mathcal{
S}*}\frac{B}{\gamma}e^{\gamma N}\left(e^{\gamma N}-1\right)
\\
\label{shs}
\mathcal{P}_\mathcal{S}^{(a)}(N) &\simeq&
\bar{\mathcal{P}}_{\mathcal{S}*}\,e^{2\gamma_*N} \, ,
\end{eqnarray}
where $\gamma=D-A$.
The quantities $\bar{\mathcal{P}}_{\mathcal{R}*}$,
$\bar{\mathcal{C}}_{\mathcal{RS}*}$ and $\bar{\mathcal{P}}_{\mathcal{S}*}$ correspond to the asymptotic limit, 
i.e. when $k\tau \to 0$, of the expressions (\ref{pra}-\ref{psa}).
 
The use of this approximation is in practice rather limited because it relies on the 
assumption that the slow-roll parameters are {\it time-independent} between Hubble crossing and  the final time. 
In most cases, this approximation, which we call {\it constant slow-roll approximation},  holds only for a few e-folds and breaks down long before the end of inflation. 

In some simple inflationary models, there exists an analytical approach  to compute analytically the evolution of 
the perturbations on super-Hubble scales. 
This is the case 
for double inflation with canonical kinetic terms (i.e. $b=0$) and potential
\beq
V(\phi,\chi) = \frac{1}{2}m_\phi^2\phi^2 + \frac{1}{2}m_\chi^2\chi^2 \, ,
\eeq
where the equations of motion for the metric perturbation 
$\Phi$ and the two scalar field perturbations can be  integrated explicitly in the slow-roll approximation and 
on super-Hubble scales \cite{Polarski:1992dq} (this can be seen as  
a particular case within a more general context discussed in \cite{Mukhanov:1997fw}).
One finds
\begin{equation}
\Phi\simeq -\frac{C_1\dot H}{H^2}+2C_3\frac{(m_\chi^2-m_\phi^2)m_\chi^2\chi^2 m_\phi^2\phi^2}
{3(m_\chi^2\chi^2+m_\phi^2\phi^2)^2}, \label{aa}
\end{equation}
\begin{equation}
\frac{\delta\phi}{\dot\phi}\simeq \frac{C_1}{H}-2C_3\frac{Hm_\chi^2\chi^2}{m_\chi^2\chi^2+m_\phi^2\phi^2},\quad
\frac{\delta\chi}{\dot\chi}\simeq \frac{C_1}{H}+2C_3\frac{Hm_\phi^2\phi^2}{
m_\chi^2\chi^2+m_\phi^2\phi^2},\label{bb}
\end{equation}
where $C_1$ and $C_3$ are time-independent constants
of integration.

The curvature and isocurvature perturbations during inflation are respectively 
\beq
\mathcal{R}=\Phi+H\, \frac{\dot\chi\, \delta\chi+\dot\phi\, \delta\phi}{\dot\chi^2+\dot\phi^2}, 
\quad
\mathcal{S}=H\, \frac{\dot\phi\, \delta\chi-\dot\chi\, \delta\phi}{\dot\chi^2+\dot\phi^2}.
\eeq
By plugging the solutions (\ref{aa}-\ref{bb}) into the above expressions, one  obtains the explicit 
evolution of the adiabatic and isocurvature perturbations, knowing that the background evolution
is given by
\begin{equation}
\chi=2M_P\sqrt{s} \sin \alpha ,\quad  \phi=2M_P\sqrt{s}
\cos \alpha, \quad s=s_0\frac{(\sin \alpha)^{2/(R^2-1)}}{ (\cos \alpha)^{2R^2/(R^2-1)}}
\label{3}
\end{equation}
where 
$s=-\ln (a/a_e)$ is the number of e-folds between a given instant 
and the end of inflation, and $R\equiv {m_\chi/ m_\phi}$.

Note that the isocurvature perturbation $\mathcal{S}$ that we have defined {\it during inflation}, following 
other works, is proportional but does not coincide with the isocurvature perturbation $S_{\rm rad}$ defined 
{during the radiation era}. In the scenario discussed in \cite{Langlois:1999dw}, where the heavy field 
$\chi$ decays into dark matter, the isocurvature perturbation $S_{\rm rad}=\delta_{\rm cdm}-(3/4)\delta_{\rm \gamma}$
is related to the perturbations during inflation via the relation 
\beq
S_{\rm rad}=-\frac{4}{3}m_\chi^2 C_3=-\frac{2}{3}\frac{m_\chi^2}{H}\left(\frac{\delta\chi}{\dot\chi}-
\frac{\delta\phi}{\dot\phi}\right).
\eeq

Another method to calculate the final curvature perturbations is the so-called
$\delta N$ formalism \cite{Starobinsky,Sasaki:1995aw,Sasaki:1998ug}. In practice however, this method 
requires the expression of the number of e-folds as a function of the initial values 
of the scalar fields and except in a few simple cases where this expression can be determined
analytically, a numerical approach is also needed in the general case.  Moreover, the approach 
we have adopted allows to follow not only the evolution of the curvature perturbation but also that of the isocurvature perturbation. This is 
important if some isocurvature perturbations  survive after the end of inflation.
Their evolution then depends on the details of the processes which occur at the end 
 inflation and after, in particular the reheating (or preheating) processes, which goes
beyond the scope of this study.

\section{Numerical analysis}

In Section \ref{inside}, we studied  the spectra and correlations
of the perturbations in the vicinity of the Hubble crossing
and we obtained analytic approximations (\ref{pra})-(\ref{psa}).
The aim of the present section is to confront these expressions
with the result of a numerical integration of the equations
of motion (\ref{eom1d})-(\ref{eom2d}). We would also like to
study numerically the super-Hubble dynamics of the perturbations, which
is often the only way to calculate the inflationary observables such as
the spectral index $n_s$ with a precision required  by the
present and forthcoming observations.

\subsection{Numerical procedure}
\label{nupro}

Our numerical procedure, similar to that of  \cite{Tsujikawa:2002qx}, is the following. 
In order to take into account the statistical independence of the 
adiabatic and isocurvature perturbations deep inside the Hubble radius, we integrate eqs.\ (\ref{eom1d})-(\ref{eom2d})
twice: first with the initial value of $\qo$ corresponding to
the Minkowski-like vacuum and $\delta s=0$, then with the 
initial value of $\ds$ corresponding to
the Minkowski-like vacuum and $\qo=0$. Unless stated  otherwise,
we impose the initial conditions 8 efolds before the Hubble crossing.
The initial conditions also include the slow-roll for the background
fields.
The evolution proceeds along a background trajectory which provides
a sufficient number of efolds before the end of inflation
(50-60, depending on the model).
We identify the end of inflation, at which we terminate the evolution,
when $\epsilon=1$.
As the outcome of the first (second) run, we obtain the curvature
and entropy perturbations, $\mathcal{R}_1$ and $\mathcal{S}_1$
($\mathcal{R}_2$ and $\mathcal{S}_2$). We then calculate the spectra and
correlations as:
\begin{eqnarray}
\mathcal{P}_\mathcal{R} &=& \frac{k^3}{2\pi^2} \left( |\mathcal{R}_1|^2+|\mathcal{R}_2|^2\right) \\
\mathcal{P}_\mathcal{S} &=& \frac{k^3}{2\pi^2} \left( |\mathcal{S}_1|^2+|\mathcal{S}_2|^2\right) \\
\mathcal{C}_\mathcal{RS} &=& \frac{k^3}{2\pi^2} \left( \mathcal{R}_1^\dagger\mathcal{S}_1+\mathcal{R}_2^\dagger\mathcal{S}_2 \right) \, .
\end{eqnarray} 
We shall sometimes describe the correlations using the relative correlation coefficient:
\beq
\tilde{\cal C} = \frac{|\mathcal{C}_\mathcal{RS}|}{\sqrt{\mathcal{P}_\mathcal{R}\mathcal{P}_\mathcal{S}}}
\label{defrc}
\eeq
The value of $\tilde{\cal C}$ lies between 0 and 1, and it indicates to what extent the final curvature
perturbations result from the interactions with the isocurvature perturbations.

\subsection{Examples of inflationary models}
\label{exex}

\begin{figure}
\begin{center}
\hspace{-2.8cm}
\includegraphics*[height=6cm]{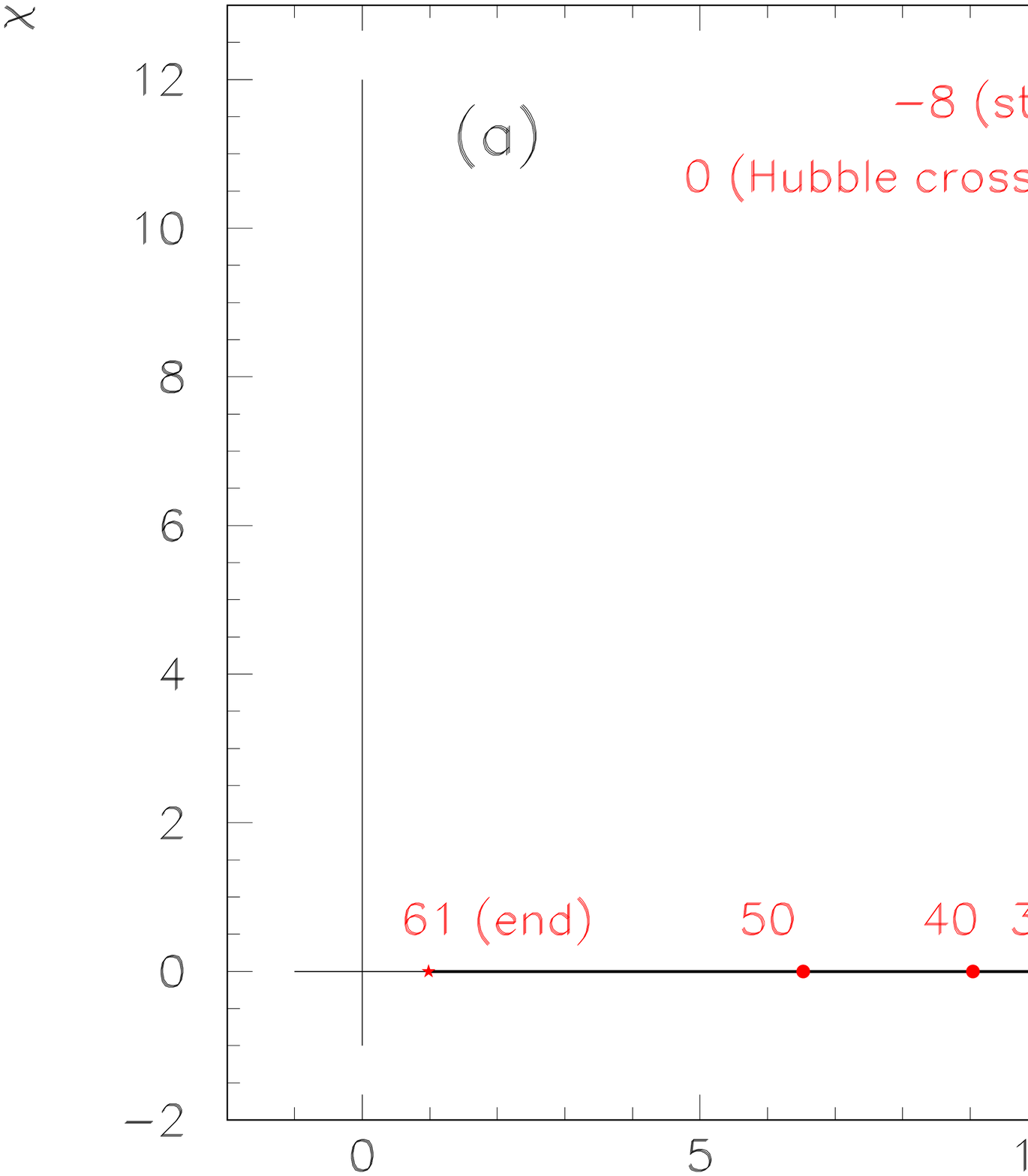}
\hspace{0.9cm}
\includegraphics*[height=6cm]{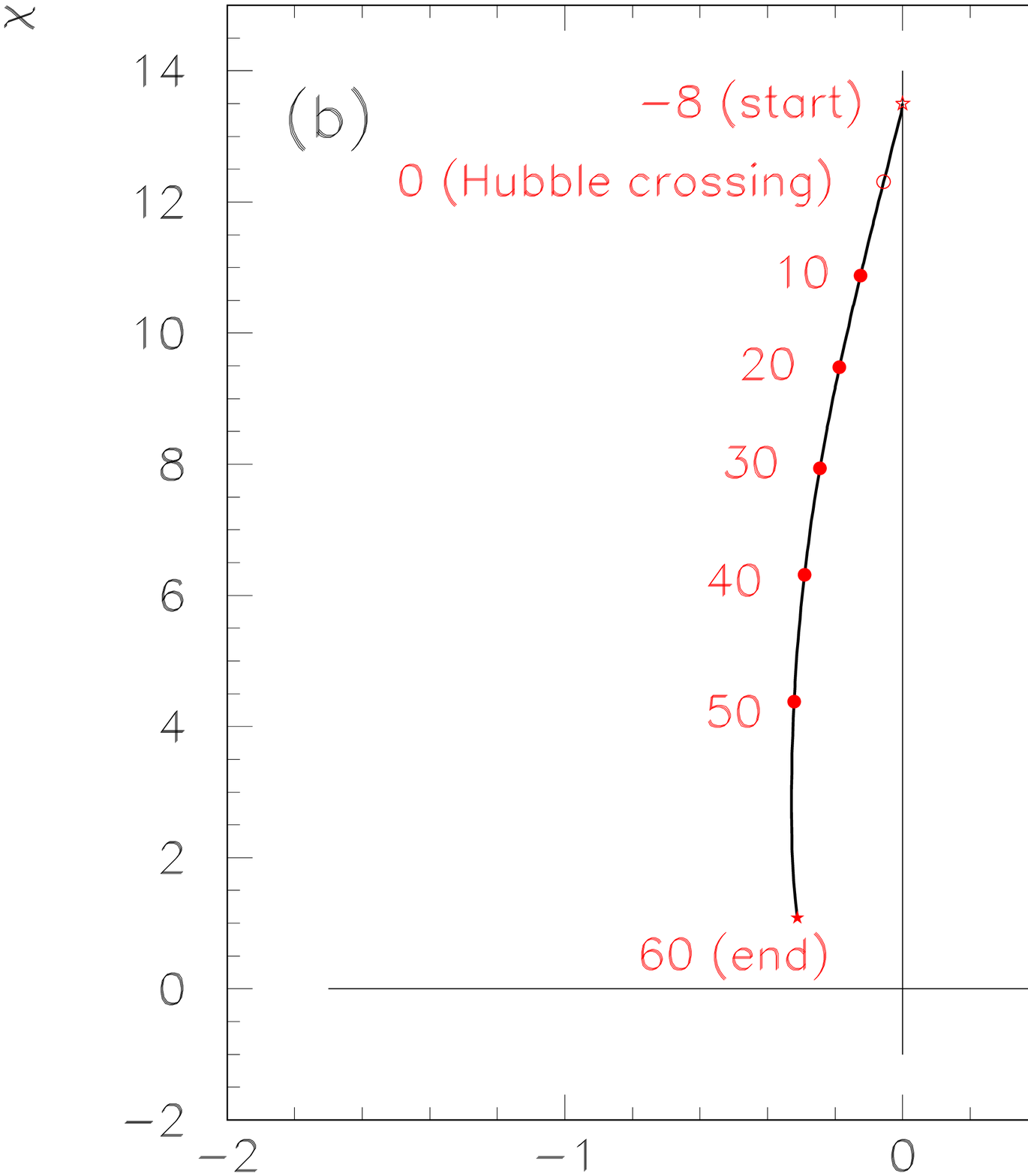}
\hspace{0.9cm}
\includegraphics*[height=6cm]{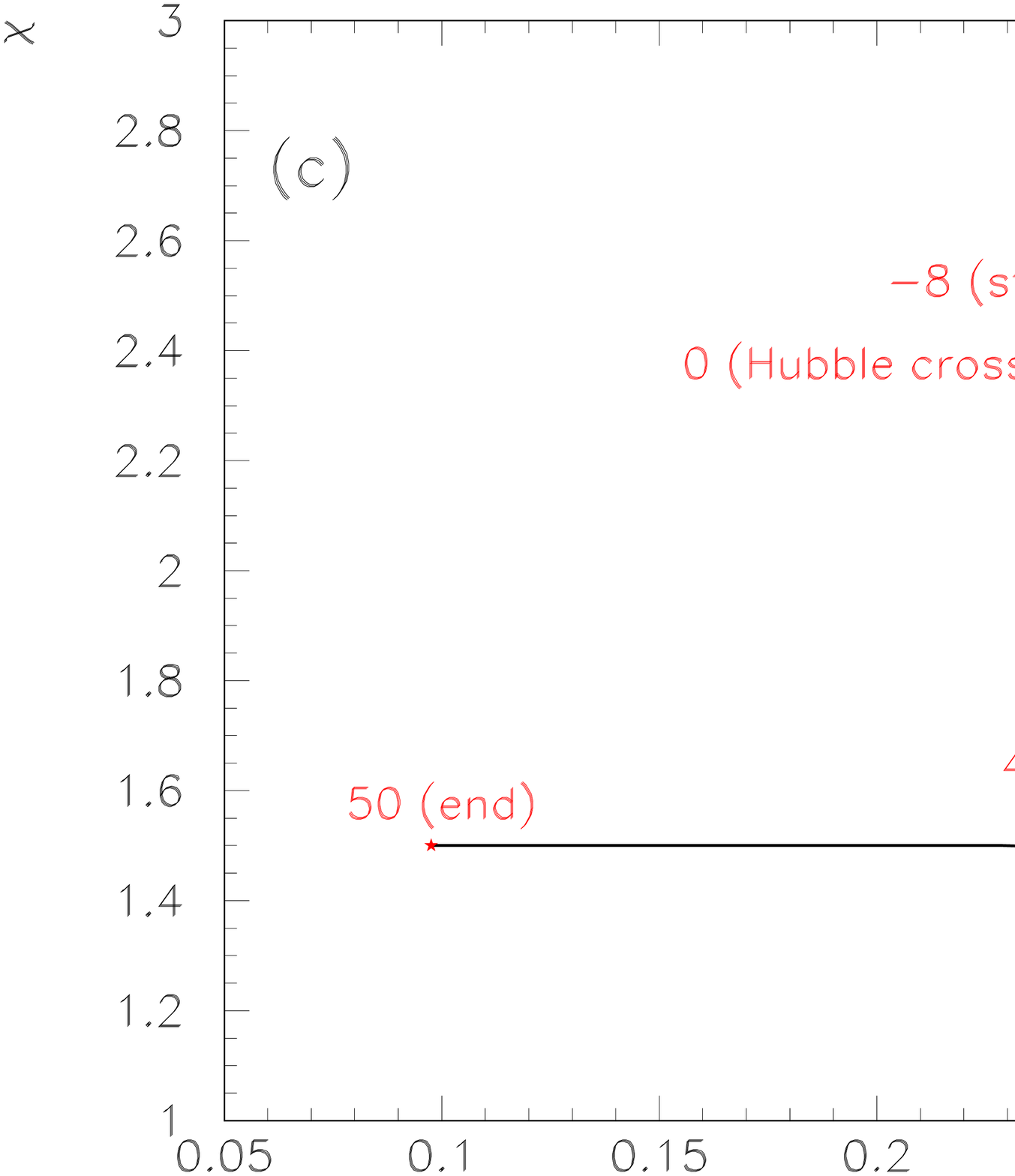}
\caption{\em Examples of classical inflationary trajectories for double 
inflation with canonical kinetic terms (left), double inflation with 
non-canonical kinetic terms (center) and roulette inflation (right).
The details of the models are described in Section \ref{exex}.
Subsequent
tens of efolds are indicated along the curves.\label{figcla}}
\end{center}
\end{figure}

There is an enormous number of examples of inflationary models. Here we
restrict our analysis to just three cases described below.
We will use these examples to check the analytical results of the preceding Sections
and to illustrate some generic features in the evolution of adiabatic and isocurvature 
perturbations.
 
\subsubsection{Double inflation with canonical kinetic terms}
\label{dickt}

Double inflation (with $b=0$) is certainly the most thoroughly studied example
of multi-field inflation. It employs the potential
\beq
\label{doublepot}
V(\phi,\chi) = \frac{1}{2}m_\phi^2\phi^2 + \frac{1}{2}m_\chi^2\chi^2 \, .
\eeq
In order to make definite calculations, we set $7 m_\phi = m_\chi$ and
we choose the initial conditions $\phi_i=\chi_i$ (later we shall also
comment on the case $\phi_i=50\chi_i$), 8 efolds before the
scale we consider  leaves the Hubble radius, after which inflation goes on for about
$\sim60$ efolds. The classical trajectory in  field space is shown
in Figure \ref{figcla}. In our example, the trajectory is strongly bent roughly  35th efolds
after the Hubble crossing.
As we shall see, it is at this moment when the adiabatic perturbations
are strongly fed by the isocurvature ones.

\subsubsection{Double inflation with non-canonical kinetic terms}

In order to concentrate just on the effects due to the non-canonical nature 
of the kinetic terms, we consider a very simple generalization of the previous example 
by taking $b(\phi)=-\phi/M_P$ and $m_\phi=m_\chi$ in (\ref{doublepot}).
We choose the initial conditions so that $\phi_i=0$. Then it is
almost exlusively the field $\chi$ which slides down to the minimum
of the potential during inflation, but due to non-canonicality
of the kinetic terms, the interaction of $\chi$ with $\phi$
drives the latter slightly away from zero.
The classical trajectory in the field space is shown
in Figure \ref{figcla}.

\subsubsection{Roulette inflation}

Recently, inflation in the large volume compactification scheme
in the type IIB string theory model has been investigated
in \cite{Bond:2006nc} (see also \cite{Conlon:2005jm}). In our notation,
this model can be effectively described by
\beq
b(\phi) = b_0 - \frac{1}{3} \ln \left(\frac{\phi}{M_P}\right)
\eeq
and
\beq
V(\phi,\chi) = V_0 + V_1 \sqrt{\psi(\phi)}e^{-2\beta_1\psi(\phi)}+V_2\psi(\phi) e^{-\beta_1\psi(\phi)}\cos(\beta_2\chi) \, ,
\label{potbkpv}
\eeq
where 
$\psi(\phi)=(\phi/M_P)^{4/3}$ and
$b_0$, $V_i$, $\beta_i$ are functions of the parameters of
the underlying string model. A generic feature of the potential
(\ref{potbkpv}) is that it has an infinite number of minima arranged
periodically in $\chi$ and a plateau for large values of $\phi$,
admitting a large variety of inflationary trajectories, which may
end at different minima even if they originate from neighboring
points in the field space -- hence the model has been dubbed
{\em roulette inflation}. In this work, we adopt the parameter set
no.\ 1 (in Planck units: $b_0 = -11$; 
$V_0 = 9.0 \times 10^{-14}$;
$V_1 = 3.2 \times 10^{-4}$;
$V_2 = 1.1 \times 10^{-5}$; and
$\beta_1 = 9.4 \times 10^5$;
$\beta_2 = 2\pi/3$)
from \cite{Bond:2006nc} and choose the particular inflationary
trajectory shown in Figure \ref{figcla}. For this trajectory,
the factor $b_\phi M_P$ is rather large, of the order $10^3$, but
the effect of the non-canonical kinetic terms is strongly suppressed
by a very small value of $\epsilon$ on the plateau of the potential.
The smallness of $\epsilon$ also suppresses the energy scale of
inflation and one needs a smaller number of efolds than in the
models described above. For definiteness, we assumed that there
are $\sim50$ efolds between the moment that the scale of interest
crosses the Hubble radius and the end of inflation.

\subsection{Numerical results for the perturbations}
\label{num}

\begin{figure}[t]
\begin{center}
\hspace{-2.8cm}
\includegraphics*[height=9cm]{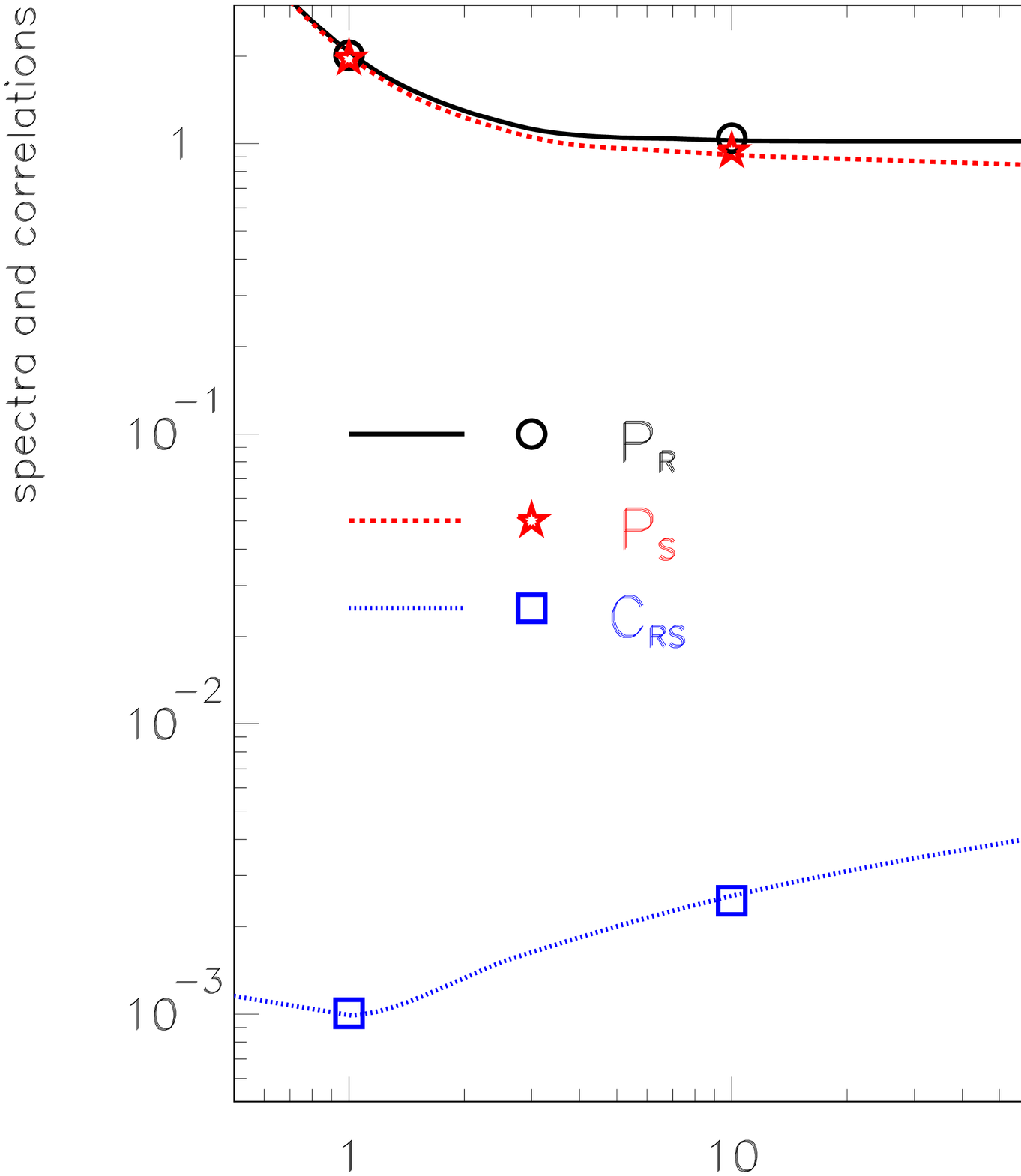}
\hspace{1.5cm}
\includegraphics*[height=9cm]{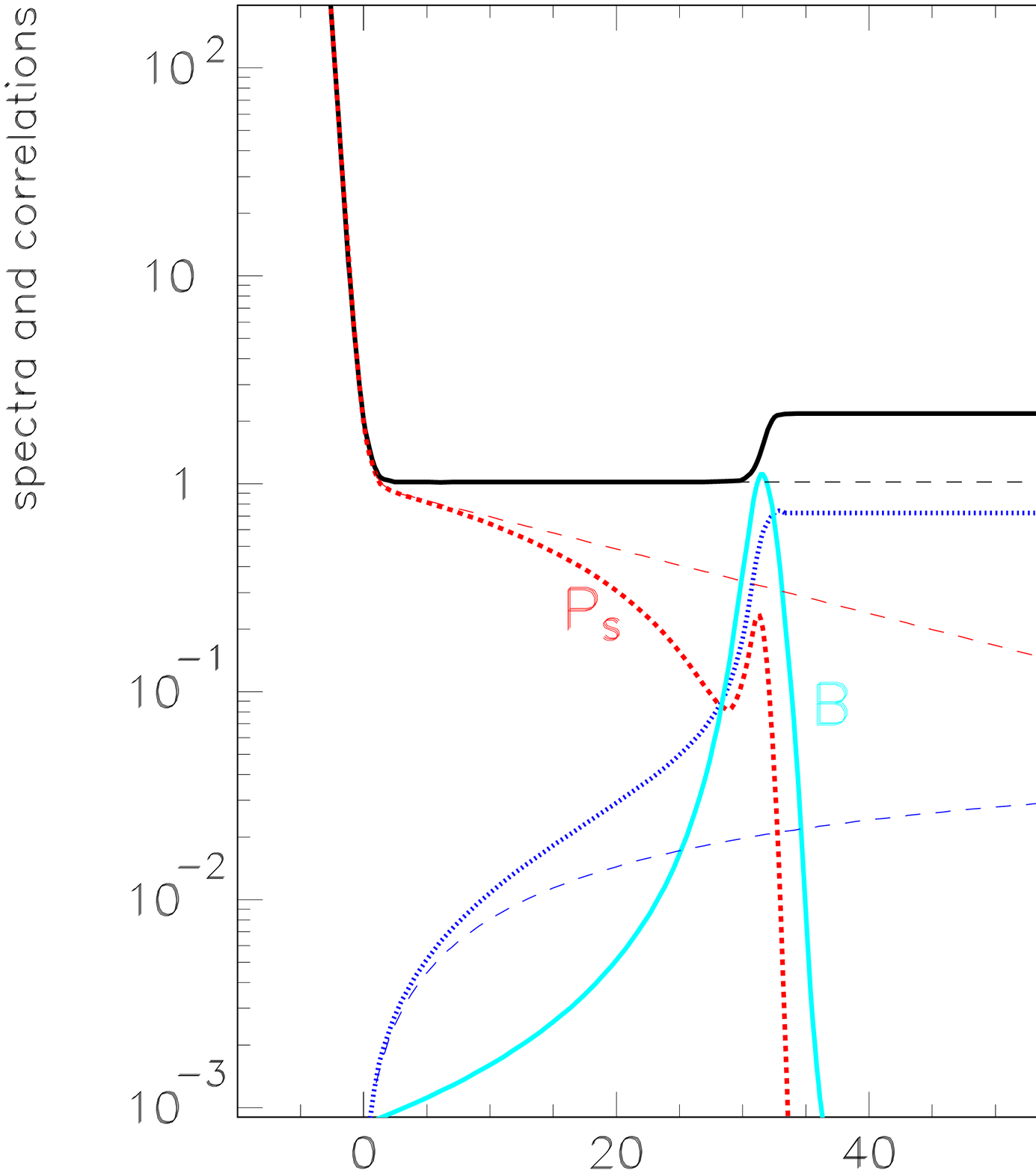}
\caption{\em 
Predictions for the spectra and correlations of the perturbations
in double inflation with canonical kinetic terms. 
Thick lines show the numerical results for ${\cal P}_{\cal R}$, 
${\cal P}_{\cal S}$ and ${\cal C}_{\cal{RS}}$ or $\tilde{\cal C}$
normalized to the single-field result (\ref{sifi}), respectively.
Circles, stars and squares indicate the predictions of
eqs.\ (\ref{pra}), (\ref{pca}) and (\ref{psa}), respectively.
Thin dashed lines indicate the predictions
of eqs.\ (\ref{shr}), (\ref{shc}) and (\ref{shs}), respectively.
The coupling $B$ between the curvature
and isocurvature perturbations is also shown.
\label{figbw}}
\end{center}
\end{figure}

\begin{figure}[t]
\begin{center}
\hspace{-2.8cm}
\includegraphics*[height=9cm]{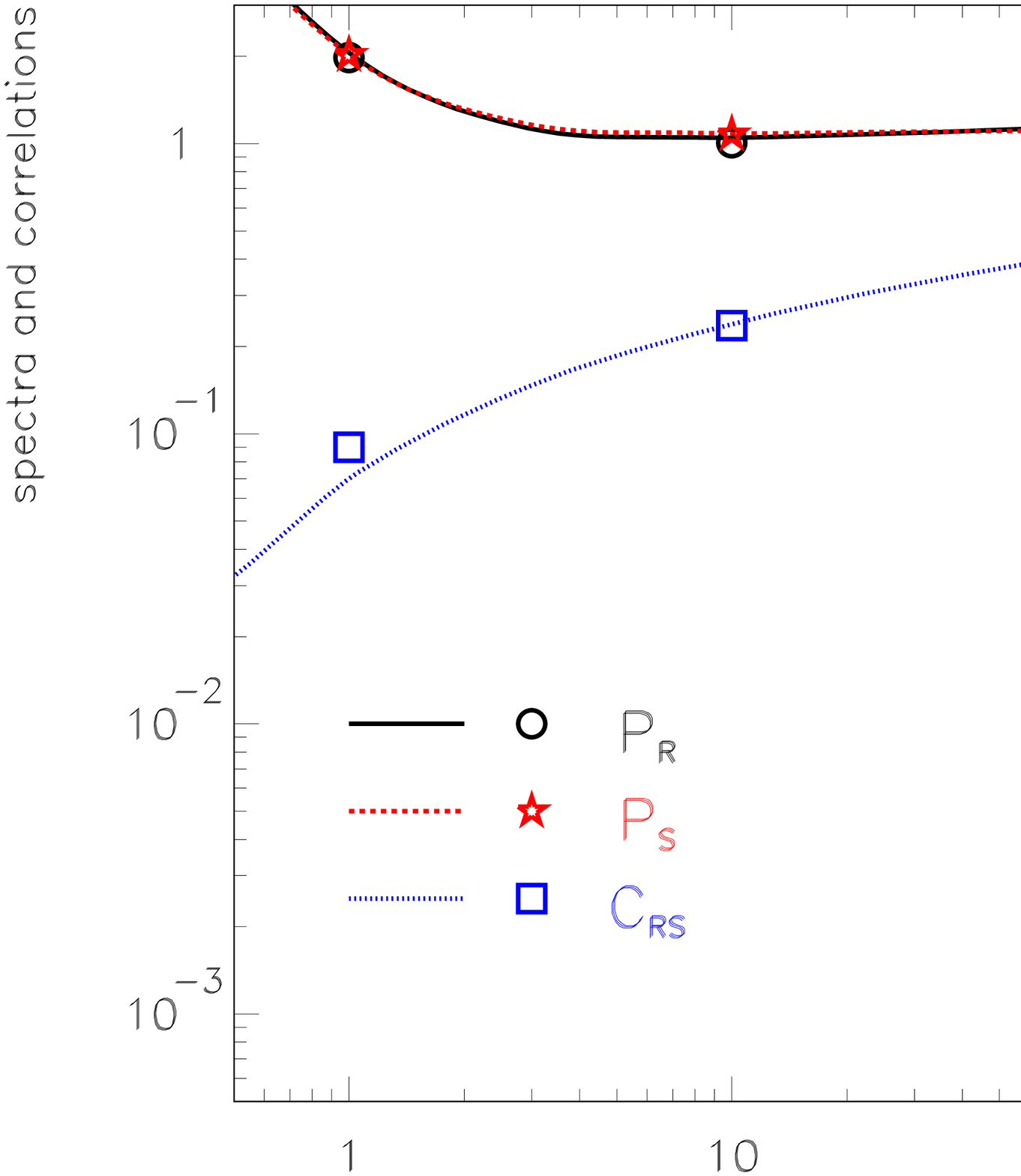}
\hspace{1.5cm}
\includegraphics*[height=9cm]{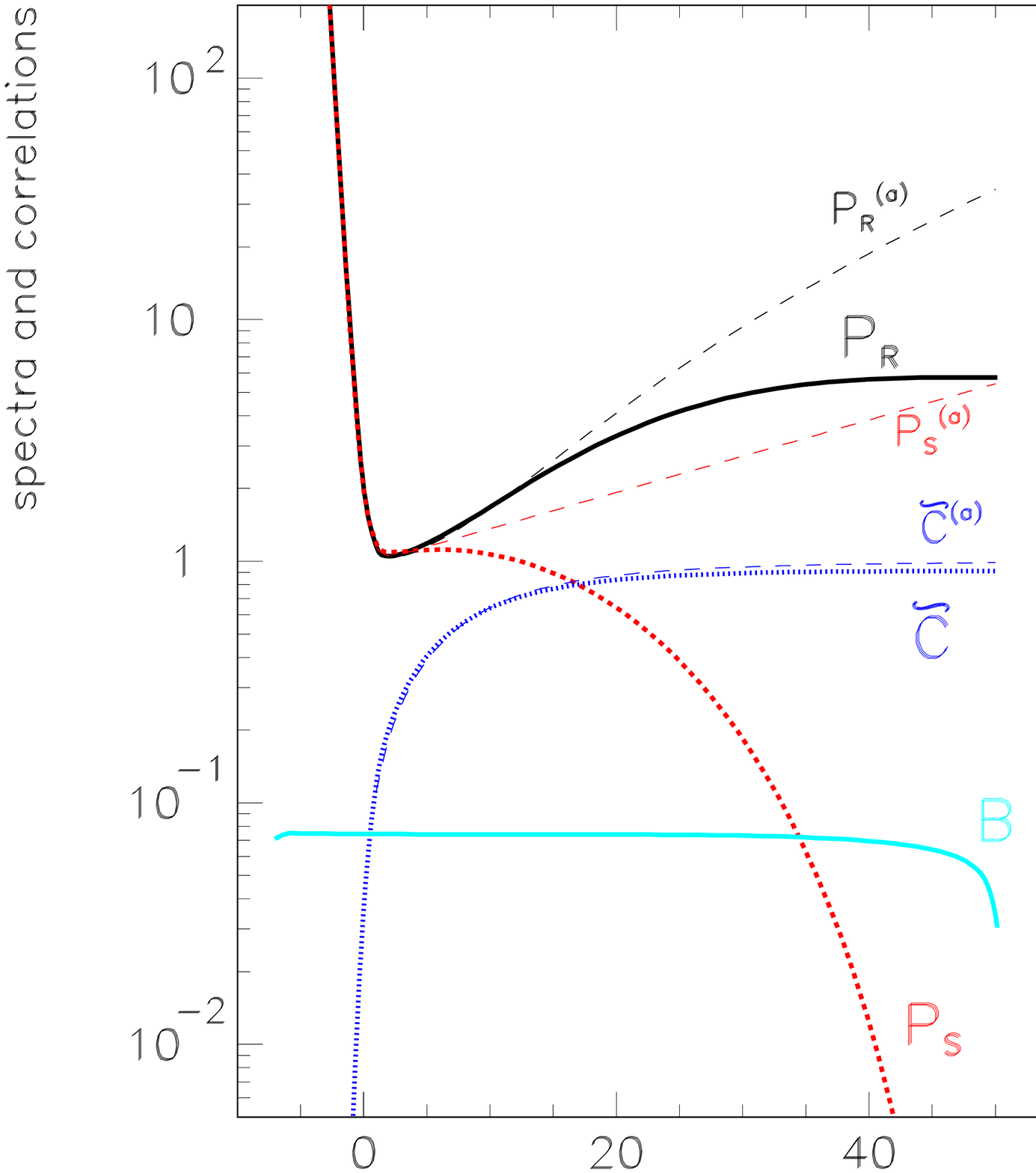}
\caption{\em 
Predictions for the spectra and correlations of the perturbations
in roulette inflation. 
Thick lines show the numerical results for ${\cal P}_{\cal R}$, 
${\cal P}_{\cal S}$ and ${\cal C}_{\cal{RS}}$ or $\tilde{\cal C}$
normalized to the single-field result (\ref{sifi}), respectively.
Circles, stars and squares indicate the predictions of
eqs.\ (\ref{pra}), (\ref{pca}) and (\ref{psa}), respectively.
Thin dashed lines indicate the predictions
of eqs.\ (\ref{shr}), (\ref{shc}) and (\ref{shs}), respectively.
The coupling $B$ between the curvature
and isocurvature perturbations is also shown.
\label{figby}}
\end{center}
\end{figure}

For the three inflationary models described in Section \ref{exex},
we performed the numerical analysis, as described in Section
\ref{nupro}. Here, we discuss the outcome for the spectra and
correlations in Figures \ref{figbw}-\ref{figby}.
In the right panel of each Figure we plot the evolution
of ${\cal P}_{\cal R}$ and ${\cal P}_{\cal S}$, 
normalized to the single-field
result ${\cal P}_{\cal R}^\mathrm{sf}$ given in eq.\ (\ref{sifi}),
as well as the evolution of the correlation coefficient $\tilde{\cal C}$
defined in (\ref{defrc}) and the parameter $B$,
defined in (\ref{bpar}), which is the coupling between the
isocurvature and the curvature perturbations.
These quantities are plotted as functions of the number
of efolds $N$ after Hubble crossing.
Left panels of each Figures \ref{figbw}-\ref{figby}
are basically close-ups of the right ones to the vicinity
of the Hubble crossing.
There,
we plot the evolution
of ${\cal P}_{\cal R}$, ${\cal P}_{\cal S}$
and ${\cal C}_{\cal{RS}}$, normalized to the single-field
result ${\cal P}_{\cal R}^\mathrm{sf}$ given in eq.\ (\ref{sifi}).
These are shown as functions of the variable $(k/aH)^{-1}$
which allows us to compare directly the numerical results with
the predictions of the eqs.\ (\ref{pra})-(\ref{psa}).
Note that in the leading order in the slow-roll parameters
$\ln(aH/k)=N$, hence the logarithmic scale in the left panels
directly corresponds to the linear scale in the right panels. 
In Figures \ref{figbw}-\ref{figby}, we also plot the evolution of the spectra, denoted by the 
superscript $(a)$, when one assumes the constant slow-roll approximation after Hubble crossing, i.e. when one 
uses  eqs.\ (\ref{shr}), (\ref{shc}) and (\ref{shs}).

For completeness, we also discuss briefly the particular case 
where the generation of isocurvature modes is effectively suppressed, situation which applies to some 
of the models discussed in the literature for specific parameters and/or initial conditions.

\subsubsection{Double inflation with canonical kinetic terms}

In this example, the field $\chi$ initially
dominates the energy density of the Universe and drives the first part
of inflation, and only when it is almost at its minimum, inflation
is further driven by $\phi$.
All the slow-roll parameters are small at the
Hubble crossing, which makes eqs.\ (\ref{pra})-(\ref{psa}) 
 an
excellent approximation of the numerical solutions of the equations
of motion. Due to the smallness of 
$B=-2\eta_{\sigma s}\approx 2\mathrm{d}\theta/\mathrm{d}N$
right after the Hubble crossing, the curvature perturbations become
practically constant during the $\chi$-domination. At the transition
to $\phi$-dominated inflation $B$ becomes large, which leads to a
sizable increment in $\mathcal{P}_\mathcal{R}$ because of interaction
with the isocurvature perturbations. During $\phi$-dominated inflation
the trajectory is almost straight again, the isocurvature perturbations
decay quickly and the curvature perturbations are frozen at the value
acquired at the transition.

This model has the advantage
that the numerical results can be directly compared to
analytical ones, as the time evolution of the
perturbations can be solved without assuming
constancy of the slow-roll parameters \cite{Langlois:1999dw}, and
we find a good agreement between two approaches.

\subsubsection{Double inflation with non-canonical kinetic terms}

\begin{figure}[t]
\begin{center}
\hspace{-2.8cm}
\includegraphics*[height=9cm]{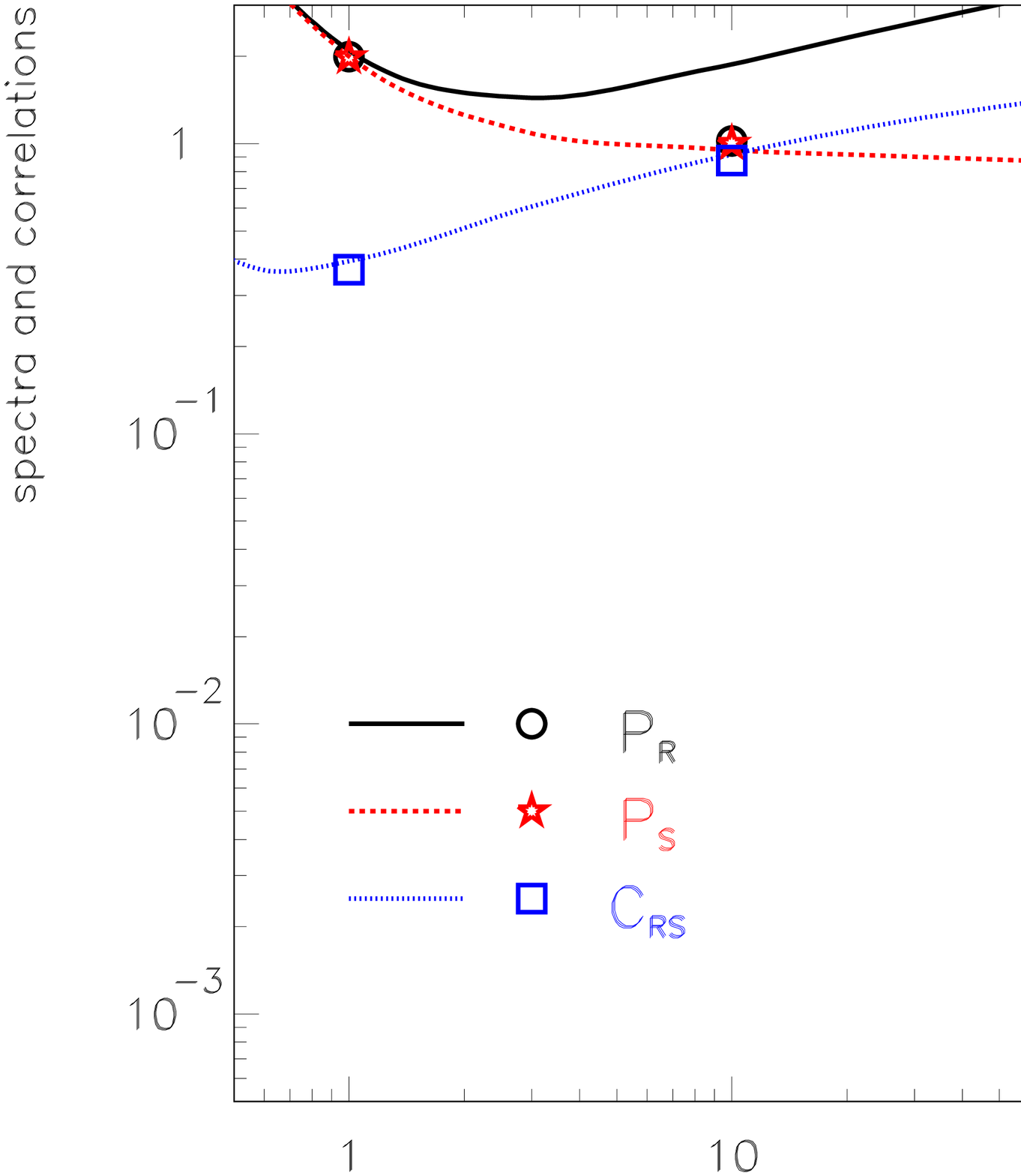}
\hspace{1.5cm}
\includegraphics*[height=9cm]{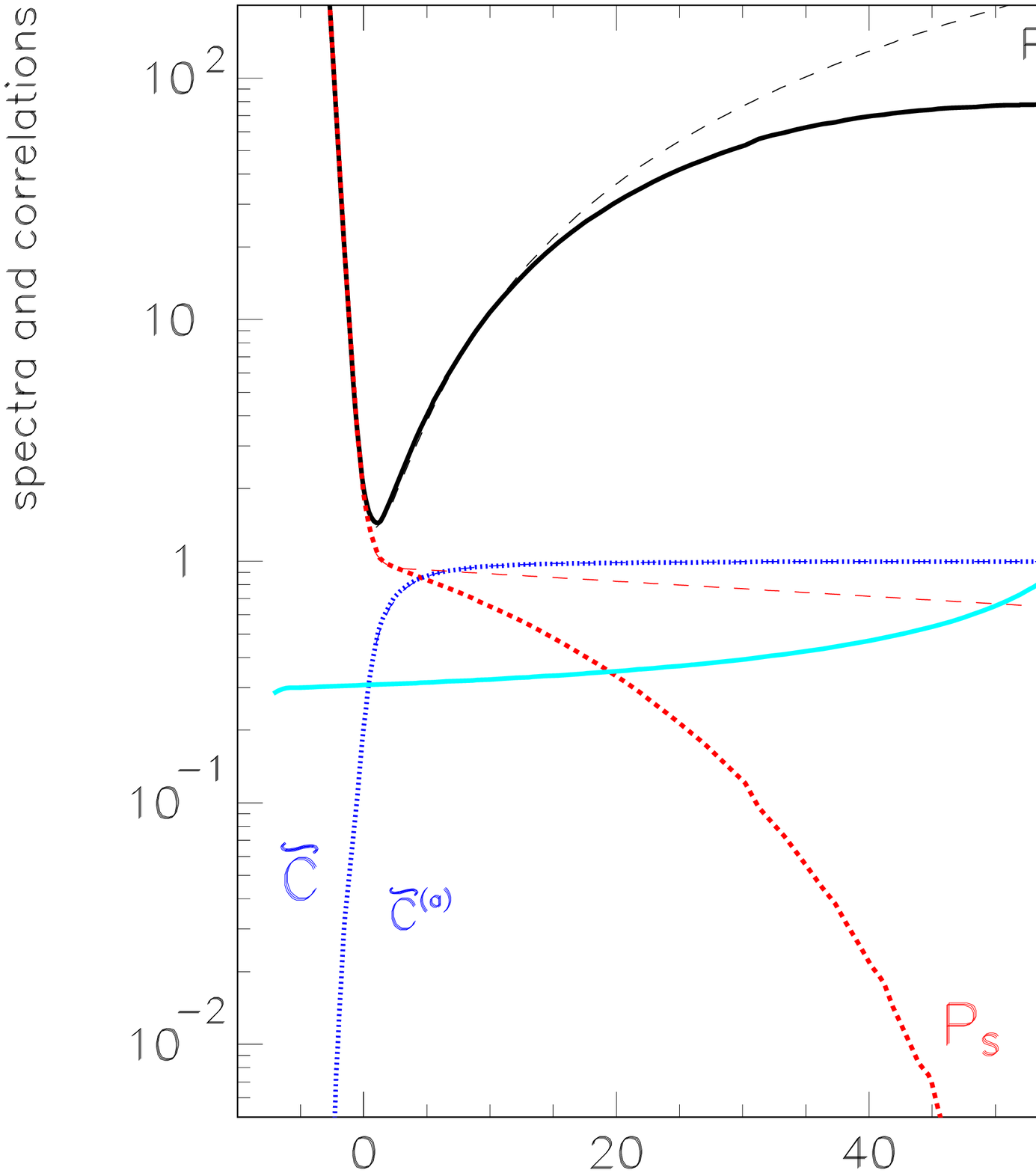}
\caption{\em 
Predictions for the spectra and correlations of the perturbations
in double inflation with non-canonical kinetic terms. 
Thick lines show the numerical results for ${\cal P}_{\cal R}$, 
${\cal P}_{\cal S}$ and ${\cal C}_{\cal{RS}}$ or $\tilde{\cal C}$
normalized to the single-field result (\ref{sifi}), respectively.
Circles, stars and squares indicate the predictions of
eqs.\ (\ref{pra}), (\ref{pca}) and (\ref{psa}), respectively.
Thin dashed lines indicate the predictions
of eqs.\ (\ref{shr}), (\ref{shc}) and (\ref{shs}), respectively.
The coupling $B$ between the curvature
and isocurvature perturbations is also shown.
\label{figbx}}
\end{center}
\end{figure}

In this example, the background trajectory is almost straight.
However, the slow-roll parameter $\epsilon$ is around $0.1$,
which makes the coupling $B$ large throughout the entire
inflationary era. Figure \ref{figbx} shows that eqs.\
(\ref{pra})-(\ref{psa})  
are a good approximation for
the spectra and correlations at the Hubble crossing, $k/aH=1$,
but it is no longer true at super-Hubble scales, 
$k/aH=0.1$ or $0.01$, because the isocurvature perturbations
already start feeding the curvature ones sizably. As a result,
the final curvature perturbations originate almost exclusively
from the interactions with the isocurvature modes, not from
the fluctuations along the inflationary trajectory, which
makes the relative correlation coefficient very close to 1.

\subsubsection{Roulette inflation}

As we already argued in Section \ref{exex}, most of the inflationary
trajectory in this example lies on the plateau of the potential
(\ref{potbkpv}), the slow-roll parameter $\epsilon$ is very small,
which makes the direct impact of the non-canonicality negligible.
The trajectory is, however, strongly curved in the field space
and the interaction between the isocurvature and curvature modes
is still important. Again, eqs.\ (\ref{pra})-(\ref{psa}) accurately predict
the spectra and correlations in the vicinity of the Hubble crossing,
with deviations on super-Hubble scales resulting from the sourcing of
the curvature perturbations by the isocurvature ones.
Eventually, most of the curvature perturbations arise through this
effect.

\subsubsection{Impact of the non-canonical terms}

In order to estimate the impact of the non canonical kinetic terms, one can separate each of the coefficients 
(\ref{cd01}-\ref{cd04})
into two parts: the terms that depend explicitly on the derivatives of non-canonical function $b$ and those which 
do not. For example, the coefficient that is directly responsible for the transfer of isocurvature modes into 
curvature modes, $C_{\sigma s}$ is decomposed into a `canonical' component $C^{\rm (c)}_{\sigma s}$ and 
a 'non-canonical' component $C^{\rm (nc)}_{\sigma s}$: 
\beq
C_{\sigma s} = C^{\rm (c)}_{\sigma s}+C^{\rm (nc)}_{\sigma s},
\eeq
with
\beq
 C^{\rm (c)}_{\sigma s}=6H\frac{V_s}{\odot}+\frac{2V_\sigma V_s}{\odot^2}+2V_{\sigma s}+ \frac{\odot V_s}{M_P^2 H},
 \quad
C^{\rm(nc)}_{\sigma s}=2b_\phi(s_\theta^3V_\sigma-c_\theta^3V_s).
\label{coedefnew}
\eeq
The other coefficients can be decomposed similarly\footnote{Note that there is some arbitrariness in this decomposition: the decomposition would be different if one expresses $V_\sigma$ in terms of $\dot\theta$ via
the background equation (\ref{dot_theta}).}.

\begin{figure}[t]
\begin{center}
\hspace{-2.8cm}
\includegraphics*[height=9cm]{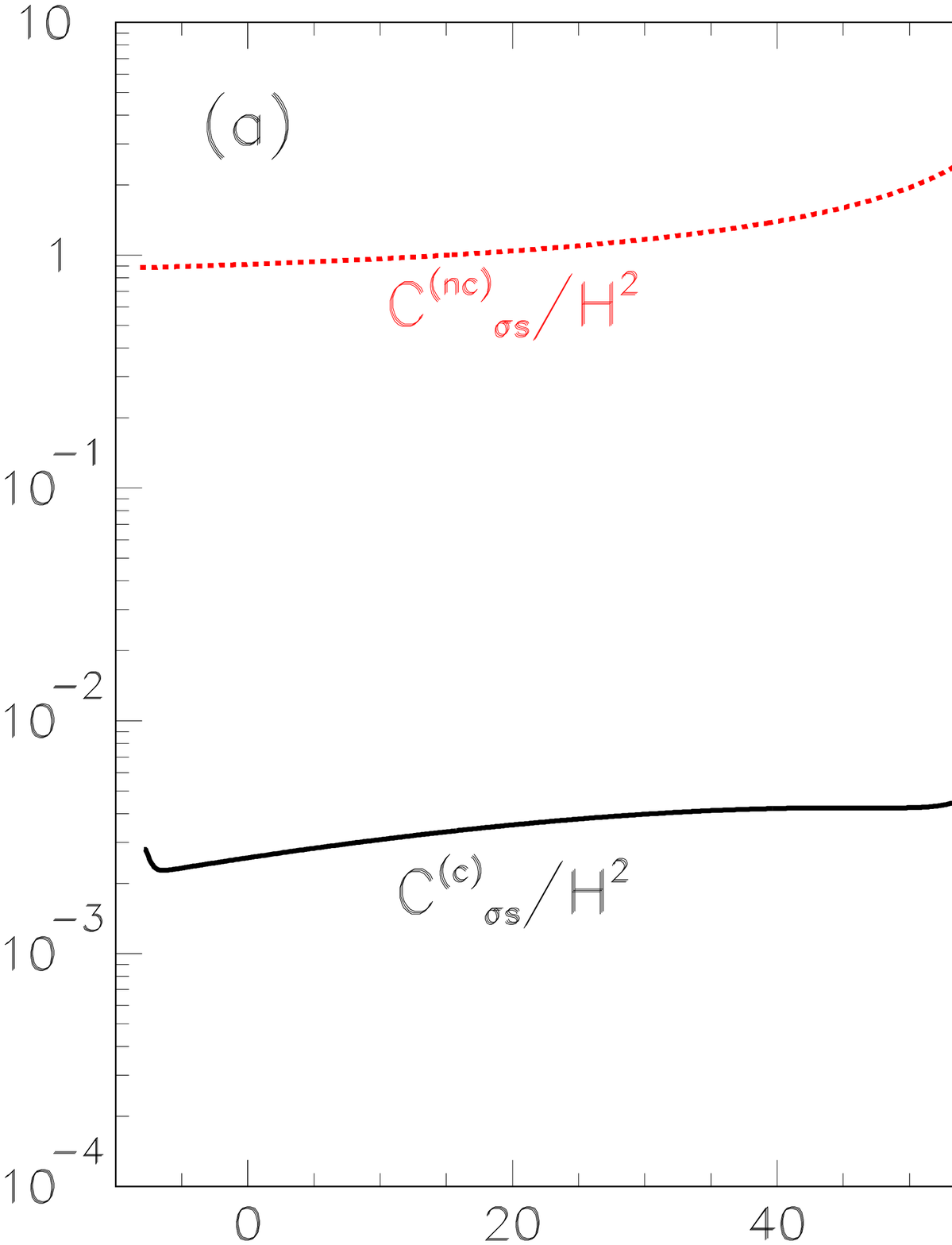}
\hspace{1.5cm}
\includegraphics*[height=9cm]{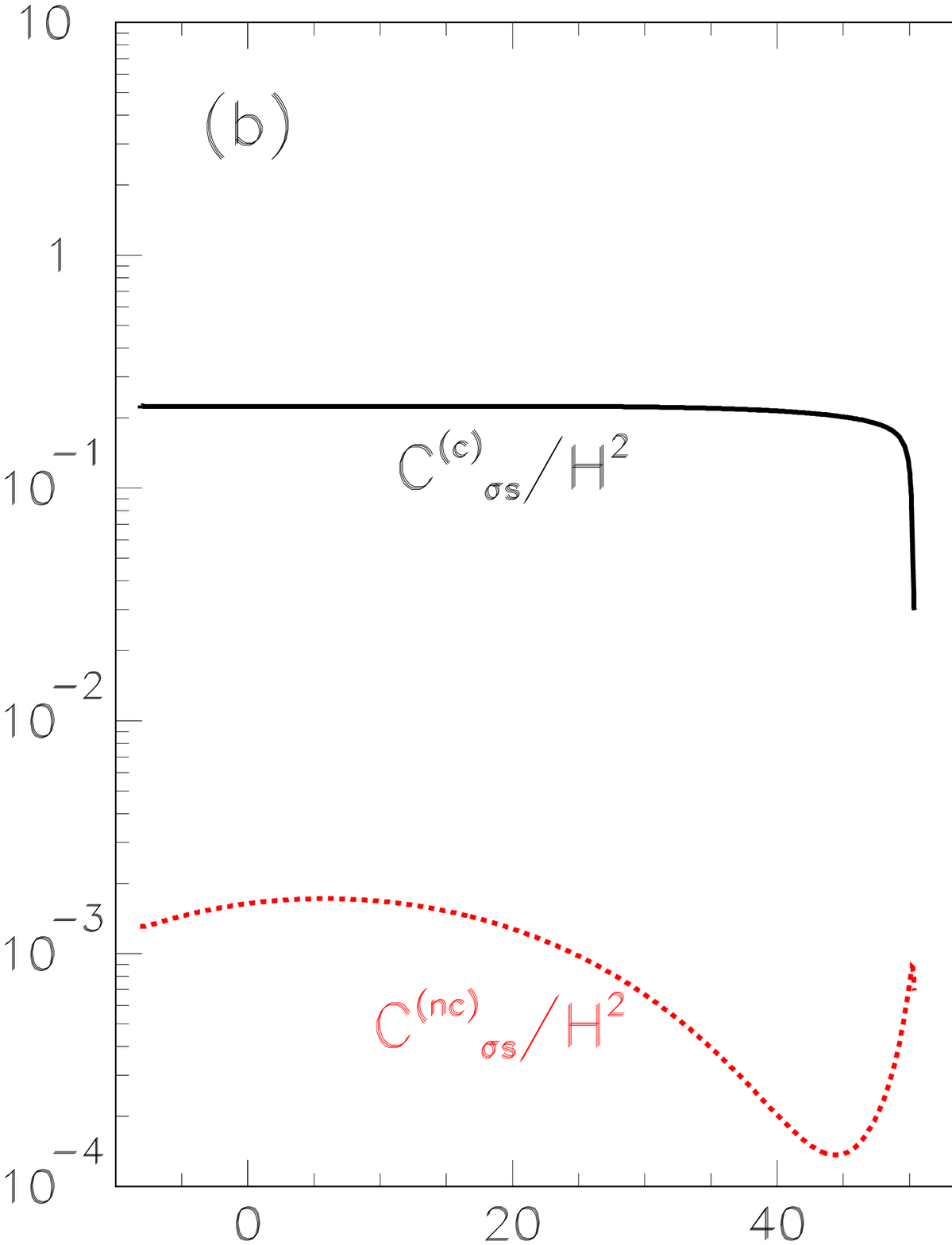}
\caption{\em 
Evolution of the coefficients $C^{\rm (c)}_{\sigma s}$
and $C^{\rm (nc)}_{\sigma s}$ defined in (\ref{coedefnew}),
parametrizing the coupling between the curvature and isocurvature
modes: (a) for double inflation with non-canonical kinetic terms
and (b) for roulette inflation.
\label{fcoeff}}
\end{center}
\end{figure}

For the two models with non-canonical kinetic terms, we have compared in Fig.~\ref{fcoeff}  the canonical and non-canonical
contributions of the various coefficients. For double inflation with non-canonical kinetic terms, the non-canonical contribution in $C_{\sigma s}$ is dominant and therefore plays a crucial role in the evolution of 
the curvature mode. For roulette inflation, as already noticed earlier, the non-canonical contributions turn out 
to be  negligible.

\subsubsection{Effectively single-field cases}
\label{esfc}

\begin{figure}[t]
\begin{center}
\includegraphics*[height=9cm]{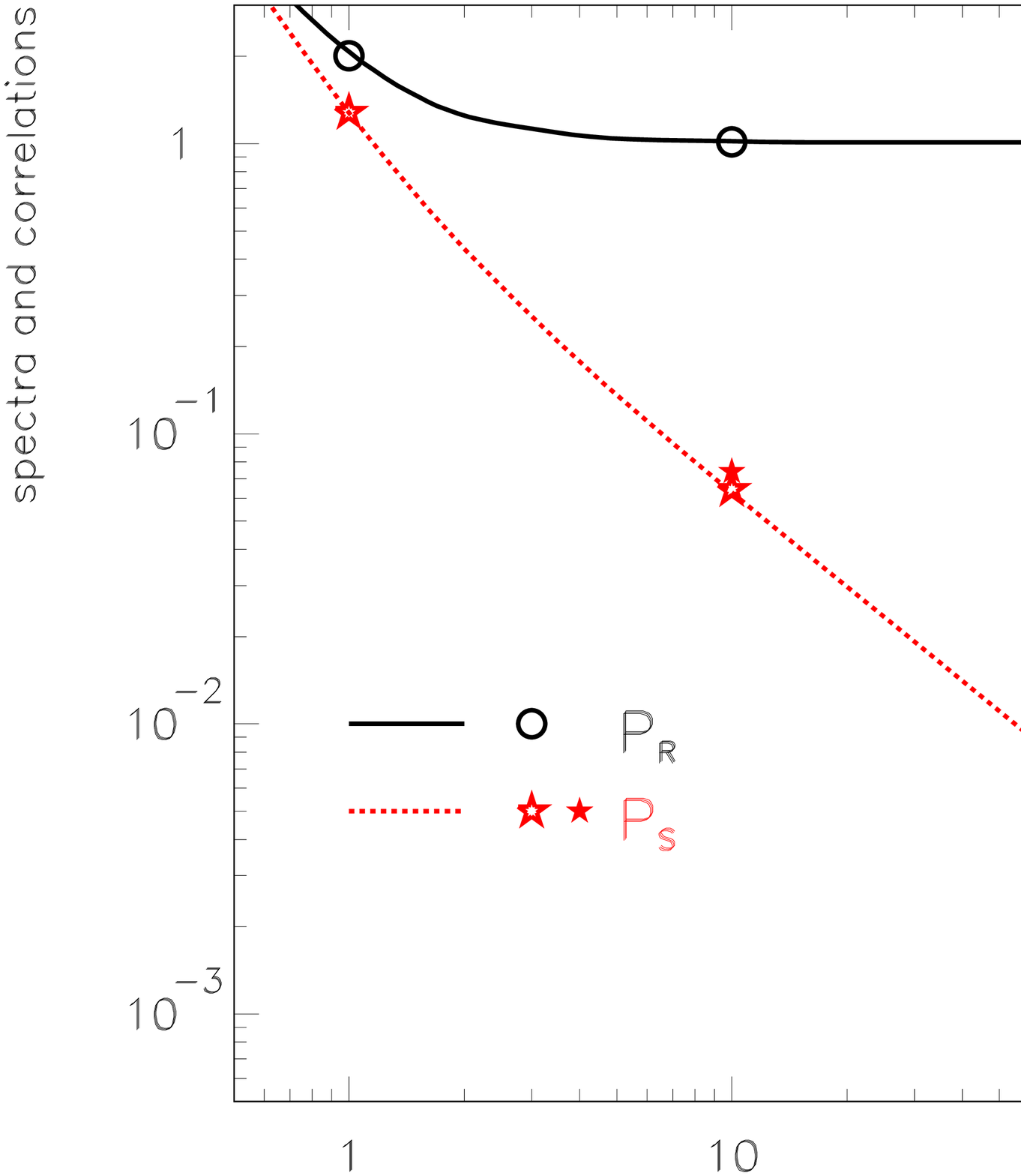}
\caption{\em 
Predictions for the spectra and correlations of the perturbations
in the effective single-field case.
The lines show the numerical results for ${\cal P}_{\cal R}$ 
and ${\cal P}_{\cal S}$,
circles correspond to prediction of eq.\ (\ref{pra}) and
stars correspond to the analytic calculation outlined in Section
\ref{esfc}. \label{fbigeta}}
\end{center}
\end{figure}

Many supergravity- or string-inspired models aim at describing
supersymmetry breaking and inflation in a unified framework.
Often, despite the presence of many scalar fields,
one can find model parameters and initial conditions such that
only for one combination of the fields a potential is sufficiently flat
to support inflation, e.g.~in pseudo-Goldstone inflation \cite{Lalak:2005hr}, 
``better racetrack'' scenario
\cite{Blanco-Pillado:2006he}, no-scale supergravity models with moduli
stabilised through D-terms \cite{Ellis:2006ar} or
in D-term uplifted supergravity of \cite{deCarlos:2007dp}.
In these works, small values of the field velocity (i.e.~$\epsilon\ll1$) 
have been ensured by setting the initial conditions in the vicinity of the
saddle point of the potential, while small curvature of the potential
along one direction has been obtained by a fine-tuning of the parameters.
It has been assumed that the isocurvature perturbations
decay fast after the Hubble radius crossing and do not affect
the curvature perturbations even though the trajectory in the field space
can have sharp turns at later stages of the inflationary evolution.
Consequently, the single field approximation (\ref{sifi}) 
has been used in these works to account for the spectra of the 
curvature perturbation.

In the two-field language developed in the present paper, 
the situation described above corresponds to
\mbox{$\eta_{ss}\simgt \mathcal{O}(1)\gg |\eta_{\sigma\sigma}|, |\eta_{\sigma s}|, \epsilon$}.
This justifies setting $\Theta_*=0$ 
in eqs.\ (\ref{bwqsigma})-(\ref{bwdeltas}), which is equivalent to the
assumption that
the curvature and isocurvature modes evolve independently.
While we can still apply the slow-roll expansion that led to eq.~(\ref{pra})
for the power spectrum of the curvature perturbations,
we now have to use the full result (\ref{solbes}) to describe
the spectrum of the isocurvature modes.
For $\eta_{ss}\simgt \mathcal{O}(1)$, the latter result decays very fast
for $k|\tau|\to 0$ and we conclude that the isocurvature modes
become irrelevant for the evolution of the curvature perturbations
soon after the Hubble radius crossing.

We checked numerically that the above conclusion applies for the models
described in \cite{Ellis:2006ar}, which are easily reduced to the
two-field case and their inflationary trajectories are curved in the 
field space. For simplicity, we would like, however, to illustrate
this point with the
model of double inflation with standard kinetic terms, described
in Section \ref{dickt}, for which we set the initial condition
\mbox{$\phi_i=50\chi_i$}.
Then the heavy field $\chi$ contributes negligibly to the potential energy
and $\eta_{ss*}\simeq 0.4$. In Figure \ref{fbigeta}, we plot the numerically
calculated spectra of the curvature and isocurvature perturbations
and compare them with the analytic approximations outlined above.
The decay of the isocurvature modes agrees with the solution
(\ref{solbes}), for which we show two cases: the small solid stars
correspond to the constant value of $\eta_{ss}$, whereas
the large empty stars show the result corresponding to adjusting
the index of the Hankel function in eq.~(\ref{solbes}) to the value of
$\eta_{ss}$ at a given instant, i.e.\ $\eta_{ss}=0.40,\,0.42,\,0.44$ for
$(k/aH)^{-1}=1,\,10,\,100$, respectively. 
With the isocurvature modes absent,
the curvature perturbations are excellently described 
by the single-field result, which justifies the use of 
the single-field approximations in the situations described above.

\subsubsection{Closing discussion}

The three examples presented here show that in multi-field
inflationary models, a large part of the curvature perturbations
can originate from interactions between the curvature and
isocurvature perturbations on super-Hubble scales, not only
from quantum fluctuations along the trajectory at the Hubble
exit. In such cases, the single-field result (\ref{sifi}) does
not provide a correct prediction either for the normalization
of the power spectrum or for its spectral index
\beq
n_s=1+\mathrm{d}\ln\mathcal{P}_\mathcal{R}/\mathrm{d}\ln k \, .
\eeq
There are techniques which allow relating the spectral index
of the curvature perturbations, $n_s$, to the spectral
indices of the entropy perturbations and the curvature-entropy
correlations through a set of consistency relations
\cite{Gordon:2000hv,Byrnes:2006fr,vanTent:2003mn},
but all these quantities separately depend on the super-Hubble
evoulution of the perturbations. Again, we find it the
most straightforward to calculate the spectral index $n_s$
for each model numerically.
In Table \ref{tablast}, we compare
naive estimate $n_s\sim1-6\epsilon_*+2\eta_{\sigma\sigma*}$ and
the predictions of eq.\ (\ref{sifi}) for the spectral index $n_s$ with 
the numerical results of Section \ref{num}. In our three examples, one can see that 
the single-field result significantly overestimates the correct spectral index. 
The discrepancies that we find follow from
the fact that the two types of perturbations experience
the slow-roll of the background fields and the curvature of
the inflationary potential in a different way. Then, if
the final curvature perturbations originate mainly from
the isocurvature ones, they inherit the features of 
the isocurvature power spectra at the Hubble crossing.

\begin{table}[t]
\begin{center}
\begin{tabular}{|c|ccc|}
\hline
$n_s$ & $1-6\epsilon_*+2\eta_{\sigma\sigma*}$ & single-field result & full result \\
\hline
double inflation (canonical) & 0.929 & 0.982 & 0.967 \\
double inflation (non-canonical) & 0.953 & 0.968 & 0.934 \\
roulette inflation & 1.017 & 1.019 & 0.932 \\
\hline
\end{tabular}
\end{center}
\caption{\em A comparison between the predictions for the spectral index
$n_s$ in the three examples of inflationary models described in
Section \ref{exex}. The third column contains result derived from
the single-field approximation (\ref{sifi}); the result of full
numerical calculations are shown in the fourth column. \label{tablast}}
\end{table}

\section{Conclusion}

In this paper, we have studied two-field inflation and extended
several previous results on curvature and isocurvature perturbations to
the case of  non-standard kinetic terms.

First, we have calculated analytically the curvature and isocurvature
spectra, as well as
the correlation, just after Hubble crossing for  two-field inflation
models, including next-to-leading order corrections in the slow-roll
approximation. Our results (\ref{pra})-(\ref{psa}) generalize those of
Byrnes and Wands, who assumed only standard kinetic terms. We have also
given a refined analytical treatment of the spectra around Hubble
crossing, which is important when the perturbations still evolve after
Hubble crossing.

Second, we have studied numerically the evolution of the curvature and isocurvature perturbations {\it after} the Hubble 
crossing. This type of analysis is important since  in multi-field inflation, in contrast with single-field inflation, 
the curvature perturbation spectrum after inflation is in general different from 
the curvature perturbation spectra at Hubble crossing because of  isocurvature perturbations,  as first
emphasized in \cite{sy}. This well-known result 
applies to multi-field inflation with either standard or non-standard kinetic terms. This effect has been studied numerically for {\it standard} kinetic terms, using the decomposition into instantaneous curvature and isocurvature, in the  analysis of \cite{Tsujikawa:2002qx}. We have done a similar analysis here for {\it non-standard} 
kinetic terms. In particular, we have compared the numerical evolution of the perturbations with an analytical approximation, which we denoted the {\it constant 
slow-roll approximation}: this approximation assumes not only that the slow-roll approximation is valid, but also that the slow-roll parameters remain almost constant during the subsequent evolution where isocurvature perturbations are significant. This approximation has been used in \cite{DiMarco:2005nq} to compute the ``final'' spectra in the context of 
non-standard kinetic terms. 
In most cases, this approximation is however not very realistic as we clearly show in our numerical study. 
We have also estimated  the impact of non-standard kinetic terms on the coupled evolution of the curvature and isocurvature modes.

There has been a recent interest in constructing inflationary models in the context of string theory. These models 
naturally lead to scalar fields with non-standard kinetic terms, to which our analysis can apply. In this work, we have studied  a very recent model, called `roulette'' inflation, and computed quantitatively the final curvature spectrum, 
thus showing that the influence of isocurvature modes on super-Hubble scales turns out to be very important.
Note that the authors of \cite{Bond:2006nc} 
computed the final curvature spectrum by
using a ``single inflaton approximation''. They stress however that
isocurvature effects ``could produce big effects'', which we indeed
confirm in our analysis.

As a message of caution for the readers who are not familiar with the effect of isocurvature modes in
 multi-field inflation, we have also computed the spectral index for curvature perturbations in a few models and contrasted 
it with the value that one would naively obtain by using the curvature perturbation at Hubble radius like in single 
field inflation. 

Finally, an interesting question, which goes beyond the scope of this paper, would be to investigate in which circumstances 
these multi-field models could produce isocurvature perturbations, after inflation and the reheating phase. 
These ``primordial'' isocurvature perturbations are today severely constrained by CMB data.

\vskip 1cm
{\bf Acknowledgements}
We would like to thank V.~Mukhanov for stimulating discussions. D.L.~would like to thank the 
Institute of Theoretical Physics of Warsaw for their warm hospitaly and for their financial support 
via a ``Marie Curie Host Fellowship for Transfer of Knowledge'' , project
MTKD-CT-2005-029466. 
Z.L.  was   partially supported by TOK project  MTKD-CT-2005-029466, by the
EC 6th Framework Programme MRTN-CT-2006-035863, and by the  grant MEiN  1
P03D 014 26.  S.P.  was partially supported by TOK project
MTKD-CT-2005-029466  and by the  grant  MEiN 1 P03B 099 29.
The work of K.T.~is partially supported by the US Department of Energy.
K.T.~would also like to acknowledge support from the Foundation for Polish Science through its programme START.

\end{document}